\documentclass{aa}  
\usepackage{graphicx}
\usepackage{txfonts}
\begin{document}
   \title{Absorption Properties and Evolution of Active Galactic Nuclei}

   \author{G. Hasinger
          \inst{1}
          \inst{2}
          }

   \offprints{G. Hasinger}

   \institute{Max-Planck-Institut f\"ur extraterrestrische Physik, 
              Gie\ss enbachstr. 1, D-85741 Garching, Germany\\
              \email{ghasinger@mpe.mpg.de} 
         \and
              Institute for Astronomy, 2680 Woodlawn Drive, Honolulu, 
             Hawaii 96822, USA\\
             }

   \date{Received ; accepted}

 
  \abstract
   { }
   {Intrinsic absorption is a fundamental physical property to understand 
   the evolution of active galactic nuclei (AGN). Here a sample of 
   1290 AGN, selected in the 2--10 keV band from different flux-limited surveys
   with very high optical identification completeness is studied.}
   {The AGN are grouped into two classes, unabsorbed (type-1) and absorbed 
   (type-2), depending on their optical spectroscopic classification and X--ray  
   absorption properties, using hardness ratios. Utilizing the optical to X--ray 
   flux ratios, a rough correction for the $\sim8\%$ 
   redshift incompleteness still present in the sample is applied. Then the 
   fraction of absorbed sources is determined as a function of X-ray luminosity 
   and redshift.}
   {A strong decrease of the absorbed fraction with X-ray luminosity is found.
   This can be represented by an almost linear decrease from $\sim80\%$ to
   $\sim20\%$ in the luminosity range log L$_X$=42--46 and is consistent with
   similar derivations in the optical and MIR bands. Several methods
   are used to study a possible evolution of the absorption fraction. A 
   significant increase of the absorbed fraction with redshift is found,
   which can be described by a power law with a slope 
   $\sim(1+z)^{0.62\pm0.11}$, saturating at a redshift of z$\sim$2.
   A simple power law fit $\sim(1+z)^{0.48\pm0.08}$ over the whole redshift 
   iis also marginally consistent with the data.} 
   {The variation of the AGN absorption with luminosity and redshift is
    described with higher statistical accuracy and smaller systematic errors
    than previous results. The findings have important consequences for the 
    broader context of AGN and galaxy co-evolution. Here it is proposed 
    that the cosmic downsizing in the AGN population is due to two different
    feeding mechanisms: a fast process of merger driven accretion at high
    luminosities and high redshifts versus a slow process of gas accretion
    from gravitational instabilities in galactic disks rebuilding around
    pre-formed bulges and black holes.}


   \maketitle
%

\section{Introduction}

In recent years it has become obvious that supermassive black holes at the 
centers of galaxies must play an important role in the evolution of galaxies. 
There is a strong correlation between the black hole mass and global properties 
of its host galaxy spheroid, like the bulge luminosity (Kormendy \& Richstone,
\cite{Kormendy1995}; Magorrian et al., \cite{Magorrian1998}) and the stellar 
velocity dispersion, i.e. the M$_{BH}$-$\sigma$ relation (Ferrarese \& Merritt 
\cite{Ferrarese2000}; Gebhardt et al. \cite{Gebhardt2000}). At the same time, 
evidence is mounting that active galactic nuclei (AGN) and galaxies in general 
undergo very similar evolution patterns. The peaks of AGN activity and star 
formation occur in the same redshift range (z = 1.5-2) and there is a similar 
dramatic decline towards low redshift. Moreover, the mass density of local 
dormant supermassive black holes in galaxy centers is consistent with the mass 
density accreted by AGN throughout the history of the Universe (Marconi et al. 
\cite{Marconi2004}; Merloni \cite{Merloni2004}), yielding further evidence for 
a tight link between nuclear black hole activity and the growth of galaxy 
bulges. Many recent theoretical models propose that feedback from the growing 
supermassive black holes plays an important role in linking the properties and 
evolution of central black holes and their host galaxy (Silk \& Rees, 
\cite{Silk1998}; Di Matteo et al., \cite{DiMatteo2005}). The feedback from 
powerful AGN can quench star formation and inhibit the further growth of the 
most massive galaxies (Scannapieco \& Oh, \cite{Scannapieco2004}), and this 
effect is indeed required to match the results of hydrodynamical simulations 
to the observed bright end of the galaxy luminosity functions (Springel et al., 
\cite{Springel2006}; Hopkins et al., \cite{Hopkins2006}).

A strong dependence of the AGN space density evolution on X-ray luminosity has
been found, the so called "luminosity-dependent density evolution" (LDDE) with
a clear increase of the peak space density redshift with increasing X--ray
luminosity, both in the soft X-ray (0.5--2 keV) and the hard X-ray (2--10 keV)
bands (Miyaji et al., \cite{Miyaji2000}; La Franca et al.,
\cite{LaFranca2002}; Cowie et al., \cite{Cowie2003}; Ueda et al.,
\cite{Ueda2003}; Fiore et al., \cite{Fiore2003}; Hasinger et al.,
\cite{Hasinger2005}). This "AGN cosmic downsizing" evolution, which recently has
also been confirmed in the radio and optical bands (Cirasuolo et al.,
\cite{Cirasuolo2005}, Bongiorno et al., \cite{Bongiorno2007}) is very similar
to the "downsizing" of the faint galaxy population which shows a smooth decline
of their maximum luminosity with decreasing redshift for z$<$1 (Cowie et al.,
\cite{Cowie1996}). The observed "cosmic downsizing" indicates that active star
formation and black
hole growth shift to lower mass galaxies throughout the evolution of the 
Universe, which is somewhat counter-intuitive to the standard Cold Dark Matter
scenario, where large scale structure evolves hierarchically from small to
larger entities. The AGN results indicate a dramatically different evolutionary
picture for low--luminosity AGN compared to the high--luminosity QSOs. While the
rare, high--luminosity objects can form and feed very efficiently, possibly by 
multiple mergers rather early in the universe (see e.g. Li et al., \cite{Li2007}), 
with their space density declining more than two orders of
magnitude at redshifts below z=2, the bulk of the AGN has to wait much longer
to grow or to be activated, with a decline of space density by less than a
factor of 10 below a redshift of one. The late evolution of the low--luminosity
Seyfert population is very similar to that which is required to fit the
mid--infrared source counts and background (Franceschini et al.
\cite{Franceschini2002}) and also the bulk of the star formation in
the Universe (Madau \cite{Madau1996}), while the rapid evolution of powerful 
QSOs traces more closely the history of formation of massive spheroids 
(Franceschini et al. \cite{Franceschini1999}). This could indicate two different 
modes of accretion and gas supply for the black hole growth with substantially 
different accretion efficiency (see e.g. the models of Cavaliere \& Vittorini 
\cite{Cavaliere2000}; Di Matteo et al., 
\cite{DiMatteo2003}; Merloni \cite{Merloni2004}; Menci et al., \cite{Menci2004}).

A large fraction of the accretion in the Universe is obscured by intervening gas 
and dust clouds. Indeed, the spectrum of the X-ray background can very well be 
described by a combination of absorbed and unabsorbed AGN evolving through cosmic 
time (e.g. Comastri et al., \cite{Comastri1995}; Gilli et al., \cite{Gilli2007}, 
hereafter GCH07). However, one of the major uncertainties in the study of AGN 
evolution and black hole growth is the unknown distribution of absorbing column 
densities and its dependence on AGN luminosity and on cosmic time. Indeed, studies 
of local Seyfert 2 galaxies have shown that a large fraction of these
($\sim$40\%) is Compton thick (Risaliti et al., \cite{Risaliti1999}) and thus 
practically absent in X--ray surveys. Therefore, the luminosity--dependent AGN 
evolution picture could be significantly biased by systematic selection effects 
(see e.g. Treister et al., \cite{Treister2004}). Ueda et al. (\cite{Ueda2003}) 
have for the first time found a significant decrease of the fraction of obscured  
AGN with increasing X--ray luminosity, and similar results have been obtained almost 
simultaneously from independent X--ray selected AGN samples by Steffen et al. 
(\cite{Steffen2003}) and Hasinger (\cite{Hasinger2004}). Recently, these trends 
have been confirmed in optically selected AGN observed in the SDSS (Simpson 
\cite{Simpson2005} and with {\em Spitzer} in the MIR (Maiolino et al., 
\cite{Maiolino2007}; Treister et al. \cite{Treister2008}), indicating a break-down 
of the standard AGN unification model (Antonucci \cite{Antonucci1993}). The possible 
evolution of the fraction of obscured sources with redshift is still a matter of 
debate in the literature. A tentative redshift dependence of the obscured fraction was 
reported by La Franca et al. (\cite{LaFranca2005}). Other authors (e.g. Ueda et al. 
\cite{Ueda2003}, GCH07) did not find a significant redshift dependence. Recently, 
Treister and Urry (\cite{Treister2006}) claimed a shallow increase of the absorbed 
fraction with redshift. The different results could indicate both 
systematic and statistical effects in the analysis.  

In this paper the best available AGN samples selected in the 2-10 keV X-ray band with the highest redshift completeness possible have been compiled, resulting in 1406 objects, of which more than 92\% have redshift information, mainly from optical spectroscopy. This allows for the so far best combination of object statistics and completeness in any hard X-ray selected AGN sample. In section 2 the hard X-ray selected AGN sample is described in detail. Section 3 discusses the classification criteria used to discriminate between type-1 (unabsorbed) and type-2 (absorbed) AGN. Section 4 presents the number counts of the different source populations and section 5 elaborates on a method to correct for the small, but significant redshift incompleteness in the sample. Section 6 is the main body of the paper and discusses the luminosity- and redshift dependence of the absorbed AGN fraction. Finally, the results are discussed in section 7 and summarized in section 8. Throughout this work a cosmology with $\Omega_m$=0.3, $\Omega_\Lambda$=0.7 and H$_0$=70 km s$^{-1}$ Mpc$^{-1}$ is used.

\begin{table*}[ht]
\begin{center}
\caption[]{The hard X--ray sample}
\begin{tabular}{@{}lccccccc@{}}
\hline
Survey       &Solid Angle& $S_{lim}$&$N_{\rm tot}$&$N_{\rm unid}$&$N_{\rm zph}$&$N_{\rm AGN1}$&$N_{\rm AGN2}$ \\
             & [deg$^2$] &[erg cm$^{-2}$ s$^{-1}$] &               &                &               && \\
\hline
HEAO1/Grossan    & 2658--22000  & (1.0--2.7)   $\cdot$ 10$^{-11}$ &   50 &   0 &   0 &  41 &   9\\
AMSS/ALSS    & $\sim$4.8--90.8  & (0.11--2.1)  $\cdot$ 10$^{-12}$ &  139 &   1 &   0 &  88 &  37\\    
HBSS         & 25.17            & 2.15 $\cdot$ 10$^{-13}$         &  67 &  2 &   0 & 42 &  20\\  
XMS          & 2.46--3.80       & (3.30--9.36) $\cdot$ 10$^{-14}$ &  144 &  13 &   0 & 111 &  25\\  
CLASXS       & 0.4              &          1.3 $\cdot$ 10$^{-14}$ &  112 &  19 &   0 &  67 &  41\\   
HELLAS2XMM   & 0.21--0.90       & (1.0--3.7)   $\cdot$ 10$^{-14}$ &  171 &  27 &   0 &  115 &  51\\
SEXSI        & 0.046--0.979     & (0.51--5.72) $\cdot$ 10$^{-14}$ &  252 &  40 &   0 & 151 &  90\\ 
LH/XMM       & 0.1056           & 5.1          $\cdot$ 10$^{-15}$ &   64 &   4 &  11 &  40 &  20\\
CDF--S       & 0.012--0.0873    & (0.44--1.49) $\cdot$ 10$^{-15}$ &  239 &   4 &  66 &  80 & 131\\
CDF--N       & 0.0075--0.0873   & (0.18--1.49) $\cdot$ 10$^{-15}$ &  284 &   8 & 111 &  99 & 141\\
\hline
Total        &                  &                                 & 1406 & 112 & 188 & 759 & 531\\        
\hline
\end{tabular}\label{tab:samp}
\end{center}
\end{table*}

\section{The hard X--ray selected AGN sample}

For the study of the AGN X--ray absorption and its cosmological 
evolution, well--defined flux--limited samples of active galactic nuclei
with a high redshift completeness have been selected in the 2--10 keV band, 
with flux limits and survey solid angles ranging over five and six orders of 
magnitude, respectively. 

A total of 1290 X--ray selected AGN were compiled from nine independent 
samples containing a total of 1406 X--ray sources selected in the 2--10 keV 
band. 1106 of these have reliable optical spectroscopic redshift identifications.
For the optically fainter sources, in particular in those samples with fainter
X--ray flux limits, it becomes increasingly difficult to obtain reliable 
spectroscopic redshifts and classifications. However, for survey fields 
with substantial multi--band photometric coverage, e.g. the GOODS North and South
areas in the CDF--N and CDF--S, respectively, it has been shown that photometric 
redshifts can be used reliably (Zheng et al. \cite{Zheng2004}, Grazian et al.
\cite{Grazian2006}). A total of 188 photometric redshifts are used
in the samples in Table~\ref{tab:samp}. The number of unidentified sources,
which do not have a reliable redshift determination, either through spectroscopy 
or through photometry, but nevertheless have optical or NIR 
counterparts, is only 112, yielding an identification fraction of 92\%.
In order to control systematic redshift incompleteness effects, crude redshifts 
have been estimated for the unidentified sources, using the X--ray to optical 
flux ratio, following Fiore et al. (\cite{Fiore2003}).  
The surveys utilized in this work are summarized in Table~\ref{tab:samp}.
Please note that the 
number of type-1 and type-2 AGN in Table~\ref{tab:samp} includes the 
photometric and crude redshifts and that other source classes like galaxies, 
stars, clusters etc. are omitted here.  

\begin{figure}[htp]
\begin{center}
\includegraphics[width=9truecm]{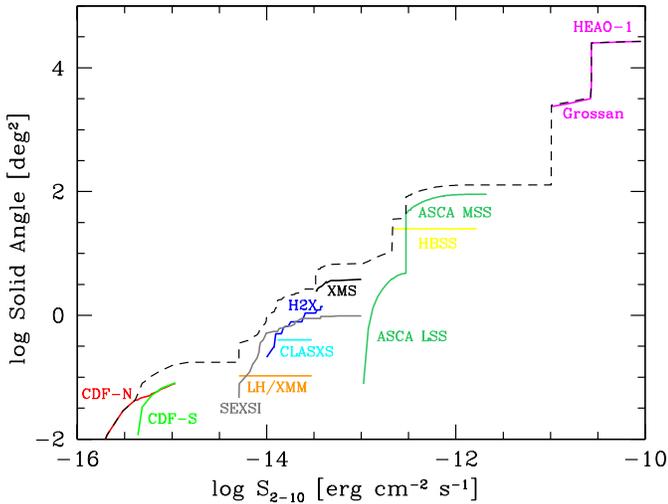}
\caption{The solid angle as a function of flux limit covered in the nine 
independent surveys utilized in this analysis.}\label{fig:Effarea}
\end{center}
\end{figure}

For several surveys a limit in X--ray flux or off--axis angle had to be chosen 
a posteriori, based on optical completeness criteria, thus maximizing the 
number of 
sources with redshifts, while simultaneously minimizing the number of 
unidentified objects. For the purpose of this paper, all X-ray fluxes 
in the different samples have been converted to observed 2-10 keV fluxes
assuming the same spectral index for all sources within each sample (see below).
While it is in principle possible to attempt a correction to {\it intrinsic},
{\it rest-frame} 2-10 keV fluxes for each source individually, using spectral
indices and hydrogen column densities determined from direct spectral fits or
hardness ratios for each source, this has not been done here for several 
reasons. This procedure introduces additional statistical errors in
the flux determination due to the uncertainty in the spectral parameters.
Most of the sources in the flux-limited samples used here contain very few
X-ray photons so that only a crude hardness ratio is available that does not
allow to determine absorption and spectral slope independently, let alone 
more complicated spectral models. While these uncertainties have a
relatively small effect on the flux observed in the source detection band,
extrapolations the source rest--frame would amplify these errors. For the 
comparison with theoretical models it is easier to fold these observational
effects into the predictions instead of trying to attempt to correct the
observations. Therefore throughout this paper all fluxes and luminosities refer
to observed quantities in the observed 2-10 keV band.  
Below we summarize our sample selection and completeness for each survey.  
Figure~\ref{fig:Effarea} gives the solid angle versus flux curve for the 
individual surveys and the total sample used.

\subsection{The {\em HEAO--1}/Grossan sample}

Shinozaki et al. (\cite{Shinozaki2006}) have selected a flux--limited 
sample of 49 AGN from the {\em HEAO--1} all--sky--survey, which has 
also been used in the Ueda et al. (\cite{Ueda2003}) analysis. For the brighter
part of the sample (28 sources with 2--10 keV fluxes larger than 
$2.7 \cdot 10^{-11}$ erg cm$^{-2}$ s$^{-1}$) they used the classical 
Piccinotti et al. (\cite{Piccinotti1982}) sample based on the {\em HEAO--1} 
A2 instrument on a solid angle of $2.7 \cdot 10^4$ deg$^2$. At fainter 
fluxes (above $10^{-11}$ erg cm$^{-2}$ s$^{-1}$) they augmented this by 
sources from the catalogue derived by Grossan (\cite{Grossan1992})
through a crosscorrelation of the {\em HEAO--1} A1 (Large--Area Counter) 
and A3 (Modulation Collimator) instruments (the {\em HEAO--1 A3 MC LASS} 
Catalog). 21 sources from the Grossan sample were selected in a sky 
area of 5105 deg$^2$ with particularly high identification completeness.
Shinozaki et al. have originally excluded the source H1443+421 (1H1448+415) 
from their low-redshift sample. However, later it turned out that the 
most likely identification for this source is the ROSAT source 
1RXS J144645.8+403510, listed as a QSO at z=0.267 in the Simbad data base
(Miyaji, 2007, priv. comm.). Including this object, the total number of 
HEAO--1 AGN used in this analysis is 50. The fluxes of the Grossan objects
have been corrected to first order for the bias related to confusion noise
by Shinozaki et al.; while these authors give fluxes for their sample based on 
individual spectral slopes, for the analysis presented here, all fluxes
have been converted assuming a power-law spectrum with a photon index
of 1.65.

\subsection{The {\em ASCA} Medium Sensitivity and Large Sky Surveys (AMSS/ALSS)}

For the {\em ASCA} Large Sky Survey and Medium Sensitivity Survey the same 
subsamples as defined by Ueda et al. (\cite{Ueda2003}) have been used in 
this analysis. 
The {\em ASCA} Large Sky Survey (LSS) is a contiguous 7 deg$^2$ strip in the 
north Galactic pole region surveyed with the ASCA GIS and SIS instruments 
(Ueda et al., \cite{Ueda1999}). Optical identifications for 34 sources above
a 2--10 keV flux limit of  $1.2 \cdot 10^{-13}$ erg cm$^{-2}$ s$^{-1}$ are given
by Akiyama et al. (\cite{Akiyama2000}) and updates in Ueda et al. 
(\cite{Ueda2003}). The {\em ASCA} Medium Sensitivity (GIS) survey is based on 
the two X--ray catalogues published by Ueda et al. (\cite{Ueda2001,Ueda2005}). 
Spectroscopic optical identifications are from Akiyama et al. (\cite{Akiyama2003}) 
for the northern subsample and from Ueda (2007, priv. comm.) for the southern
subsample. In total the {\em ASCA} Large Sky and Medium Sensitivity Surveys
contain 139 objects, of which 125 are AGN and only one source remains unidentified.

\subsection{The {\em XMM--Newton } Hard Bright Serendipitous Sample (HBSS)}

The HBSS is a subset of the larger XMM-Newton Bright Serendipitous Survey 
(Della Ceca et al. \cite{DellaCeca2004}) carried out by the XMM-Newton Survey 
Science Center (Watson et al.  \cite{Watson2001}) consortium. It is a complete 
flux-limited sample of bright X-ray sources 
($F_x>7\cdot10^{-14}~erg~cm^{-2}~s^{-1})$ at high galactic latitude ($|b|>20\deg$), 
selected in the 4.5-7.5 keV energy band. The HBSS was designed to have a flat sky 
coverage of 25.17 deg$^2$ at a fixed MOS2 count rate limit of 
$2 \cdot 10^{-3}~cts~s^{-1}$ in the 4.5-7.5 keV band. The MOS2 count rates in the 
HBSS were converted to observed fluxes in the 2-10 keV band, assuming a photon power 
law index of 1.9 (using a count rate to flux conversion factor of 
$1.077 \cdot 10^{-11}~erg~cm^{-2}~cts^{-1}$. The flux limit of the HBSS is therefore 
$2.15 \cdot 10^{-13}~erg~cm^{-2}~s^{-1}$. The HBSS sample is now almost completely 
optically identified, leaving only 2 out of 67 objects unidentified and thus yielding 
a spectroscopic completeness of $\sim 97\%$ (Caccianiga et al. \cite{Caccianiga2008}; 
Della Ceca et al. \cite{DellaCeca2008}).

\subsection{The {\em XMM--Newton } Medium--sensitivity Survey (XMS)}

The {\em XMM--Newton} Medium-sensitivity Survey (XMS, Barcons et al.
\cite{Barcons2007}) is a serendipitous X--ray survey and optical identification 
program of sources with intermediate X--ray fluxes discovered in 25 {\em XMM--Newton} 
high Galactic latitude fields covering a sky area of $\sim$3~deg$^2$. For this 
analysis the XMS--H sample, selected in the 2--10 keV band was used. 
The original Barcons et al. sample contains 159 sources with 2--10 keV fluxes 
$>3.3 \cdot 10^{-14}$ erg cm$^{-2}$ s$^{-1}$ with a spectroscopic identification
fraction of 83\% (27 unidentified sources). However, the actual spectroscopic 
completeness limit varies from field to field. In order to optimise the redshift 
completeness of the XMS, a strategy was chosen, which maximizes the sample of 
identified sources, while minimizing the number of unidentified sources. 
There are 15 fields which either are completely identified down to the survey 
flux limit, or have a small number of unidentified sources significantly brighter 
than the flux limit. For those fields, the flux limit was maintained and 
the unidentified sources enter the sample for this analysis. For the remaining 10 
fields, unidentified sources have the lowest flux in the particular subsample.  
In these cases the unidentified sources have been excluded. The flux limit for 
this field was then raised to the geometric mean between the flux of the first 
unidentified and that of the last identified source in order to avoid the 
'gerrymandering effect' introduced by defining a post--facto flux limit (see
also Hasinger et al. \cite{Hasinger2005}). This effect can be explained in
the following way: if one has a pre-defined flux limit for a (sub)survey, then
the faintest source in the sample is expected to lie somewhat above this flux
limit. If, on the other hand, the flux limit would be defined post facto 
exactly at the flux of the faintest identified source, a small bias is 
introduced. Defining the flux limit between two adjacent sources in the 
original sample minimizes this bias. Table \ref{tab:xms} summarizes the 
unidentified source statistic for the XMS fields. This way a cleaner XMS sample 
could be defined, comprising 144 objects and almost half the number of 
unidentified sources, i.e. achieving an identification fraction of 91\%. 
The survey solid angle has been corrected accordingly (see Fig. \ref{fig:Effarea}).

\begin{table}[ht]
\begin{center}
\caption[]{Identification completeness in XMS fields used in this study}
\begin{tabular}{lccccccc}
\hline
Number of unidentified Sources & 0  & 1 & 2 & 3 \\
Number of Fields     & 16 & 6 & 2 & 1 \\
\hline
\end{tabular}\label{tab:xms}
\end{center}
\end{table}

\subsection{The {\em Chandra} Large Synoptic X--ray Survey (CLASXS)}

CLASXS is a large contiguous survey consisting of nine intermediate--depth {\em Chandra} 
pointings in a $3 \times 3$ grid in the Lockman Hole ($\sim2.5\deg$ Northwest of the 
original LH deep survey, see below), in a region with the lowest Galactic neutral hydrogen 
column density (Lockman et al. \cite{Lockman1986}). A catalogue of 525 X--ray sources
discovered in CLASXS is given in Yang et al. (\cite{Yang2004}). Steffen et al. 
(\cite{Steffen2004}) have presented optical identifications and 271 spectroscopic 
redshifts for the CLASXS, yielding an overall spectroscopic completeness of 52\%. 
For the work discussed here, the sample of 319 sources detected with a signal--to--noise
ratio $SNR \ge 2.0$ in the hard band (2--8 keV) has been utilized. Down to the original 
survey flux limit of $2.1 \cdot 10^{-15}$ erg cm$^{-2}$ s$^{-1}$, there are 138 objects 
without redshifts.
In order to obtain an acceptable unidentified fraction, the original catalogue had to be 
cut at a substantially higher flux limit of $1.3 \cdot 10^{-14}$ erg cm$^{-2}$ s$^{-1}$, 
yielding a sample
of 103 sources, of which 20 remain unidentified. The redshift completeness is thus 81\%.

\subsection{HELLAS2XMM}

HELLAS2XMM (Baldi et al., \cite{Baldi2002} is a serendipitous survey based on suitable {\em XMM--Newton} 
pointings (complementary to the XMS) and building on the experience of the original 
{\em BeppoSAX} High Energy Large Area Survey (HELLAS). Fiore et al. (\cite{Fiore2003}) 
have presented optical identifications and spectroscopic redshifts for 122 sources 
selected in the 2--10 keV band in five {\em XMM--Newton} fields, covering a survey 
area of 0.9~deg$^2$. For the discrimination between absorbed and unabsorbed sources
(see below), hardness ratios from Perola et  al. (\cite{Perola2004}) were used, 
kindly provided by M. Brusa (2007, priv. comm.). 
Originally Fiore et al. published optical spectroscopic redshift
identifications for 97 of their objects, yielding a redshift completeness of $\sim 80\%$.
Later Maiolino et al. (\cite{Maiolino2006}) presented additional redshifts from VLT NIR 
spectroscopy of optically extremely faint objects for two additional sources. Both 
Fiore et al. and later Mignoli et al. (\cite{Mignoli2004}) tried to 
estimate redshifts for the remaining, optically faint unidentified sources. Fiore et 
al. employed the optical to X--ray flux ratio to obtain crude redshift information. 
Mignoli et al. used further, deep NIR imaging with ISAAC at the VLT of optically very 
faint counterparts in HELLAS2XMM. Using both the R--K colours and the morphological 
information, they estimated 'photometric' lower limit redshifts for 9 of the originally 
unidentified sources in the Fiore et al. sample, yielding also consistent redshifts 
for the two objects later identified in the Maiolino et al. NIR spectroscopy. 
In section \ref{sec:crude} the X--ray/optical flux ratio will be used to estimate 
crude redshifts for all unidentified sources in this study, but for the moment the 
crude redshifts for HELLAS2XMM are not taken into account in the sample completeness.
However, a 2--10 keV flux limit of $10^{-14}$ erg cm$^{-2}$ s$^{-1}$ was chosen, thus  
removing the faintest 4 sources, which are all unidentified. 

Recently Cocchia et al. \cite{Cocchia2007}, published photometry and spectroscopic redshifts in five additional HELLAS2XMM fields providing 59 
new redshift identifications for the sample of 110 new sources. In order to 
maximize the redshift completeness of this sample a procedure similar to the 
one used for the XMS has been applied (see above). However, because in this case
some fields were observed or identified only outside the central image region, 
an inner and
an outer off-axis cut had to be applied in addition to a flux cut. As in the 
case of the XMS, the flux and off-axis limits were chosen in the geometric mean 
between the last identified and the first unidentified object. This way flux 
limits between $1.18$ and $2.56 \cdot 10^{-14}$ erg cm$^{-2}$ s$^{-1}$ were 
chosen for the different fields and the total additional solid angle collected 
this way was $0.48~deg^2$. The sky coverage of the survey (see Cocchia et al. 
\cite{Cocchia2007}) was corrected accordingly. In this way an additional 49 
sources with only 4 unidentified objects could be added to the sample. 
This leaves in total 171 sources, of which 27 remain unidentified. The redshift
completeness in this subsample of HELLAS2XMM is thus 84\%. 

\subsection{The Serendipitous Extragalactic Source Identification {\em SEXSI} Programme}

SEXSI is a serendipitous survey based on 27 {\em Chandra} archival, high Galactic latitude
pointings followed up spectroscopically with the Keck telescopes. Eckart et al.  
(\cite{Eckart2006}) presented spectroscopic redshift identifications for 438 objects 
out of a total sample of 1034 sources selected in the 2--10 keV band, yielding 
an original identification completeness of 42\% over a survey area of $\sim2$~deg$^2$. 
For their sample Eckart et al. have utilized the {\em Chandra} sources over the
whole field--of--view (FOV) and to the full sensitivity limit of their observations.
In order to optimise the completeness of the SEXSI sample to be utilized for this 
study, the original sample was cut both in off-axis angle and in flux limit, 
following a procedure similar to that described above for the XMS. At the outset, 
the whole sample has been reduced to off--axis angles below $8.31\arcmin$, thus 
concentrating on the inner part of the {\em Chandra} FOV with higher angular 
resolution and sensitivity. Then, a limit was defined separately for each of the 
27 SEXSI pointings both in flux and in off--axis angle in a way to limit the fraction 
of unidentified sources to $<80\%$ in each field. A total of 7 out of the original 
27 fields had to be rejected, because this condition could not be fulfilled. The 
remaining fields had between 0 and 5 unidentified sources left (see Table 
\ref{tab:sexsi}). In order to avoid biases in the effective area versus solid 
angle curve introduced by this 'gerrymandering' procedure, both the flux limit 
and the off-axis limit for each pointing was defined as the geometric mean between 
the corresponding values of the last identified and the first unidentified
source in each field (sorted in flux or in off--axis angle, respectively).
In this way, a total number of 252 objects were retained for the SEXSI subsample
used in this study, of which 40 remain unidentified, yielding a spectroscopic 
completeness of $\sim84\%$ over a solid angle of $\sim1$~deg$^2$.    

\begin{table}[ht]
\begin{center}\caption[]{Identification completeness in SEXSI fields used in this study}
\begin{tabular}{lccccccc}
\hline
Unidentified Sources & 0 & 1 & 2 & 3 & 4 & 5 & not used\\
Number of Fields     & 1 & 8 & 5 & 3 & 2 & 1 & 7\\
\hline
\end{tabular}\label{tab:sexsi}
\end{center}
\end{table}

\subsection{Deep {\em XMM--Newton} survey of the Lockman Hole (LH/XMM)}
\label{sec:xmmlh}

The Lockmam Hole (LH/XMM) has been observed by {\em XMM--Newton}  
17 times during the PV, AO--1 and AO--2 phases of the mission, with total good 
exposure times in the range 680--880 ks in the PN and MOS instruments (see 
Hasinger et al. \cite{Hasinger2001,Hasinger2004}, Worsley et al. \cite 
{Worsley2004} and Brunner et al. \cite{Brunner2008} for details). Spectroscopic
optical identifications of the {\em ROSAT} sources in the LH have been presented 
by Schmidt et al. (\cite{Schmidt1998}) and Lehmann et al. (\cite{Lehmann2000}).
A catalogue from the {\em XMM--Newton} PV phase was published by Mainieri et al. 
(\cite{Mainieri2002}) and some photometric redshifts have been discussed in Fadda 
et al. (\cite{Fadda2002}). Here objects were selected from 266 sources detected in the hard band of the 637 ksec dataset 
(Brunner et al. \cite{Brunner2008}) with additional spectroscopic identifications 
obtained with the DEIMOS spectrograph on the Keck telescope in spring 2003 and 2004
by M. Schmidt and P. Henry (Szokoly al., \cite{Szokoly2008}) and also including 
a number of new photometric redshifts (Mainieri, 2007, priv. comm.). In order to 
maximize the spectroscopic/photometric completeness of the sample, objects were 
selected in two off--axis intervals with different 2--10 keV flux limits: 
$2.5 \cdot 10^{-14}$ erg cm$^{-2}$ s$^{-1}$ for off--axis angles in the range 
11.0--14.3 arcmin and $5 \cdot 10^{-15}$ erg cm$^{-2}$ s$^{-1}$ for off--axis 
angles smaller than 11 arcmin. The total number of sources in the LH/XMM 
survey is 64, with 4 objects still unidentified.

\begin{figure*}[htp]
\begin{center}
\includegraphics[width=14truecm]{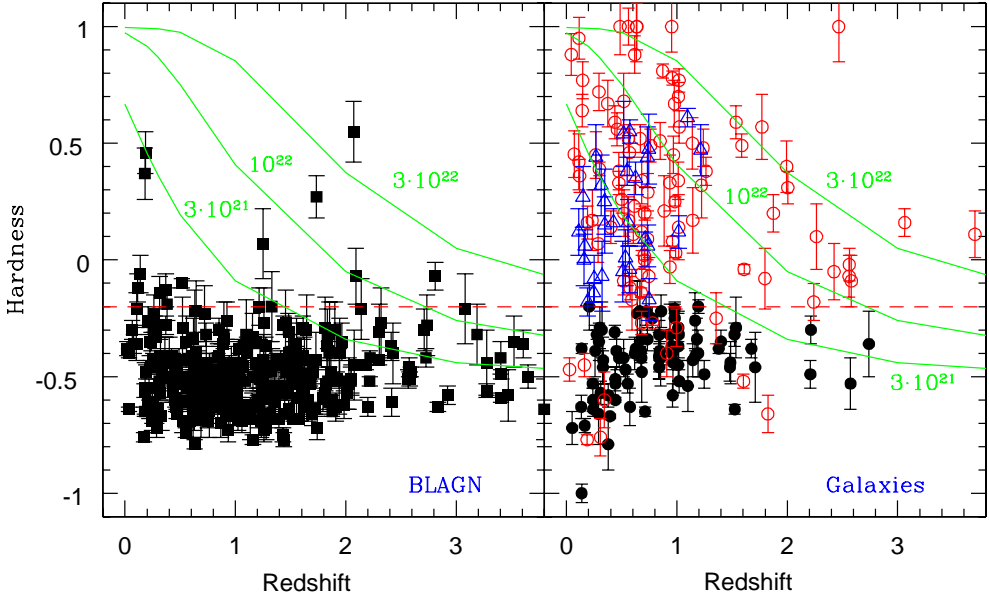}
\caption{The hardness ratio of X--ray sources selected from {\em XMM--Newton}
and {\em Chandra} surveys with reliable optical spectroscopic
classification and hardness ratio errors less equal than 0.15 as a function 
of redshift. Left: 320 objects optically classified as broad--line AGN (BLAGN). 
Right: objects without broad lines in their optical spectra. Sources classified 
as type-1 AGN by their X--ray hardness ratios (109 objects) are indicated by 
filled circles, spectroscopically classified type-2 AGN (200 objects) by open 
circles
and objects spectroscopically classified as normal galaxies by green triangles (140 
objects). The green solid lines in both diagrams are model predictions for a grid of 
power law AGN spectra with photon index 2, locally absorbed through different column 
densities, as indicated in the plot. }\label{fig:hrz}
\end{center}
\end{figure*}

\subsection{The {\em Chandra} Deep Field South (CDF--S)}\label{sec:cdfs}

For this study the catalogue of Giacconi et al. (\cite{Giacconi2002}), based 
on the 1 Ms observation of the CDF--S (Rosati et al. \cite{Rosati2002}), was used
and expanded with two additional objects from the catalogue derived 
from the same data set by Alexander et al. (\cite{Alexander2003}) which were
confused in the original Giacconi et al. catalogue, but formally fulfill the same 
detection criteria. One source (XID 527) was removed, because it was identified 
with the jet of an extended radio source already matched with another X-ray source
(Mainieri et al., \cite{Mainieri2008}). 239 sources detected significantly in 
the 2--10 keV band (signal to noise ratio $>$2.0) and within an off--axis 
angle of 10 arcmin were thus selected. The solid angle area curve given in 
Giacconi et al. (\cite{Giacconi2002}) was used, but truncated at an off--axis
angle of 10 arcmin. 

Spectroscopic identifications in the CDF--S have originally been obtained by 
Szokoly et al. (\cite{Szokoly2004}) with the FORS instruments at the ESO VLT, 
yielding a 
spectroscopic completeness of $\sim48\%$ in the hard band. Additional spectroscopic 
redshifts of CDF--S X--ray sources have been obtained with the ESO VLT as part of 
the VVDS and GOODS surveys and have been collected in the catalogue of Grazian et 
al. (\cite{Grazian2006}). In the meantime more spectroscopic redshifts have
been obtained both with VIMOS and with FORS2 at the VLT through the GOODS programme
(PI: C. Cesarsky) and the CDFS follow-up (PI: J. Bergeron). Recently also the 
spectroscopic follow--up observations 
of the Extended Chandra Deep Field South (ECDFS, see Lehmer et al., \cite{Lehmer2005})
have yielded additional identifications in the CDF--S proper, both from the VLT and 
from Keck (G. Szokoly, J. Silverman, 2007, priv. comm.). For the 239 CDFS  
sources selected for this study we have therefore the following breakdown of spectroscopic redshifts: 111 objects from Szokoly et al. (\cite{Szokoly2004}),
30 redshifts from the public GOODS data releases based on VLT FORS and VIMOS
spectroscopy (Vanzella et al., \cite{Vanzella2008}; Popesso et al., \cite{Popesso2008}), 13 and 4 objects from
the ECDFS spectroscopic follow--up using VIMOS at the VLT and DEIMOS at the 
Keck telescopes, respectively (J. Silverman, G. Szokoly et al., 2007, priv. comm.), 
7 objects from the K20 
survey (Mignoli et al. \cite{Mignoli2005}) and 3 redshifts from the VVDS (Le Fevre 
et al. \cite{LeFevre2004}). This subsample of the CDF--S has therefore a spectroscopic
completeness of 71\%. 

\begin{figure*}[htp]
\begin{center}
\includegraphics[width=9truecm]{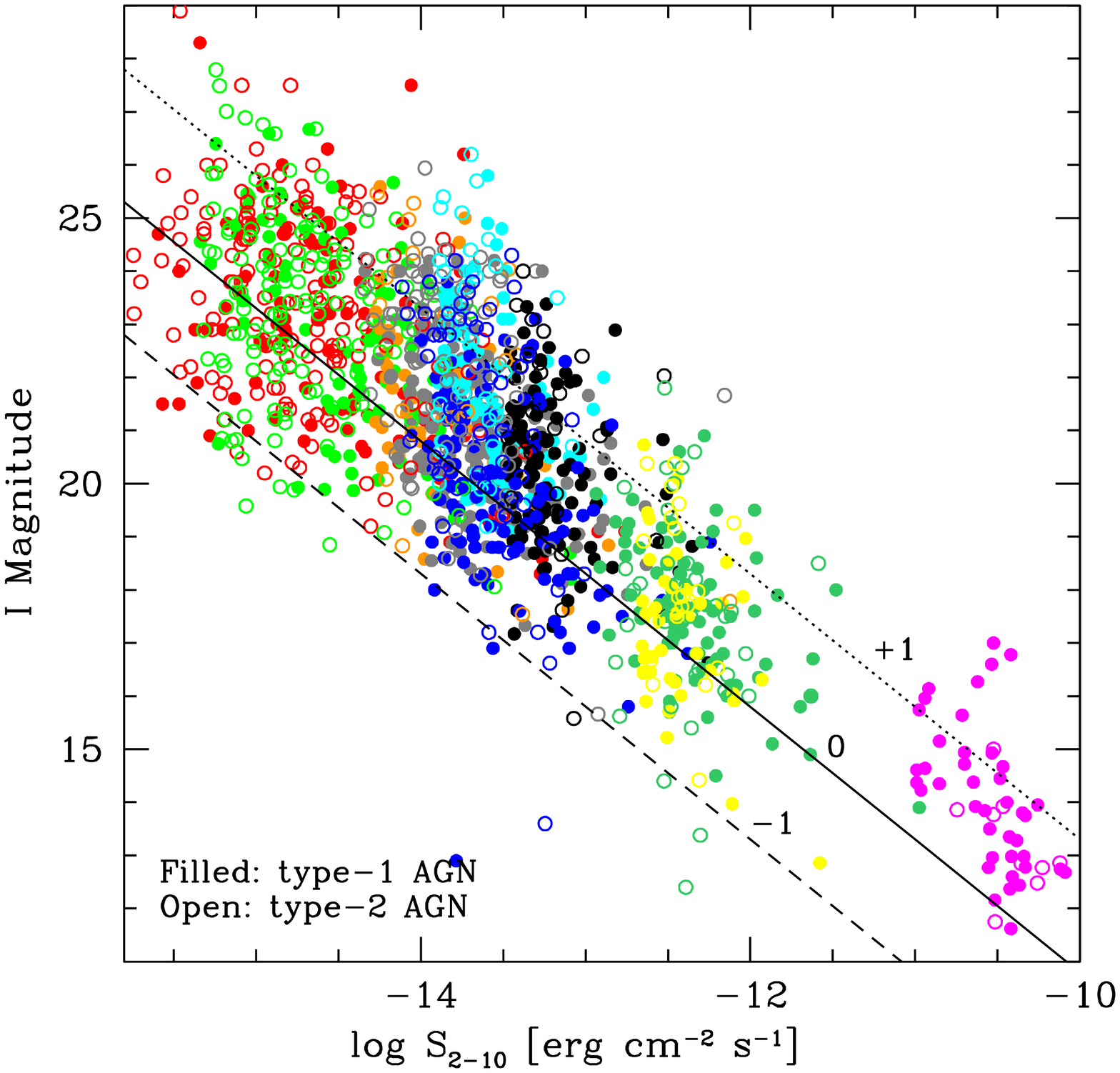}
\includegraphics[width=9truecm]{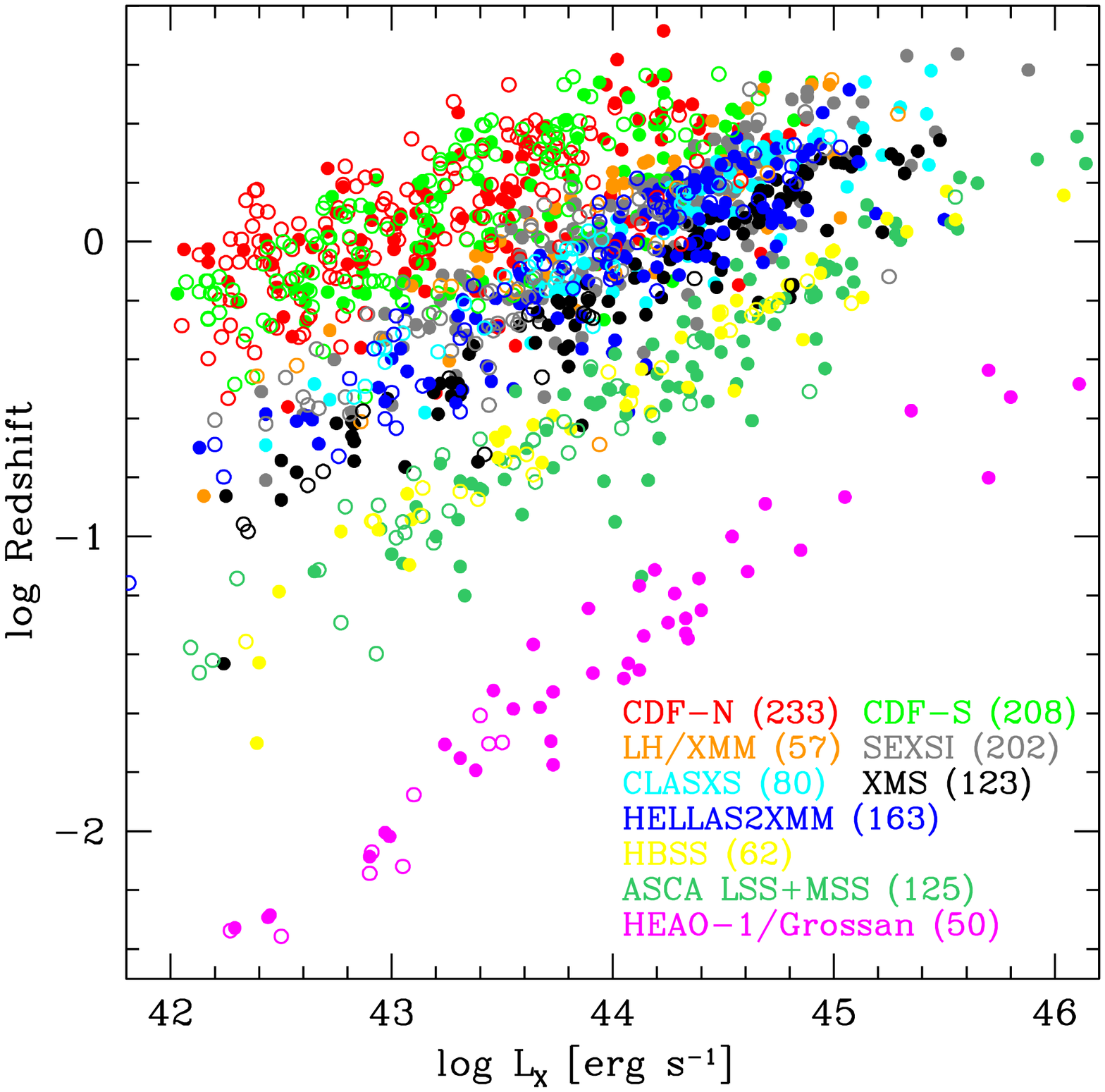}
\caption{Left: Hard X--ray (2--10 keV) sample of 1290 AGN in the I--Magnitude / 
2--10 keV flux plane. The dotted, solid and dashed diagonal lines indicate 
constant X--ray to optical flux ratios of log(f$_X$/f$_{opt}$=-1,0, and +1,
respectively. Filled symbols refer to type-1, and open circles to type-2 
AGN as defined in Section~\ref{sec:class}. Right: The same data in the 
redshift / luminosity plane. Luminosity here and throughout the paper
refers to the observed luminosity, i.e. uncorrected for intrinsic
absorption.}\label{fig:hsam}
\end{center}
\end{figure*}

The field is also included in the COMBO-17 intermediate--band survey, which 
gives very reliable photometric redshifts for optically brighter sources 
(Wolf et al., \cite{Wolf2004}). A total of 9 COMBO-17 redshifts have been used
here. The CDF--S has also been surveyed by the HST ACS as part of the GOODS 
(Giavalisco et al. \cite{Giavalisco2004}), and GEMS (Rix et al. \cite{Rix2004}) 
projects. Very deep NIR photometry has been obtained with the ISAAC camera at the 
VLT as part of the GOODS project (Mobasher et al. \cite{Mobasher2004}). Finally, the 
GOODS area has been covered in the MIR spectral range by deep Spitzer 
observations. The CDF--S therefore offers the highest quality photometric redshifts 
of faint X--ray sources, which have originally been derived by Zheng et al. 
(\cite{Zheng2004}) and Mainieri et al. (\cite{Mainieri2005}). More recently, Grazian 
et al., (\cite{Grazian2006}) have derived photometric redshifts in the GOODS CDF--S 
area including the Spitzer MIR photometry. Using the better infrared photometry, 
Brusa et al. (\cite{Brusa2008}) have obtained new, unambiguous identifications for 
two objects, which originally had two plausible counterparts in Szokoly et al. 
(\cite{Szokoly2008}): For the X--ray source with XID 201 the spectroscopic redshift 
z=0.679 in Szokoly et al. was given for the object 'b' in the {\em Chandra} error 
circle, while the Spitzer data indicate
that the object 'a' with a photometric redshift z$_p$=2.88 (Grazian et al.) is 
the correct counterpart.  A similar situation exists for the X--ray source with 
XID 218, where the spectroscopic redshift z=0.479 for object 'a' has been 
replaced with a photometric redshift z=2.27 for object 'b' in the error circle.
Several of the original Zheng et al. photometric redshifts have been superseded by the 
better Grazian et al. values, however, the overall redshift distribution for the CDF--S did 
not change significantly using the new data. For this analysis 27 photometric 
redshifts taken from Grazian et al. were used, and 33 redshifts with a photometric 
quality $Q\ge0.2$ from Zheng et al. were utilized. 
Four objects had too low a quality in their photometric redshift solution or no
data at all and thus remain unidentified, yielding an excellent overall 
redshift completeness of 98\%.

\subsection{{\em Chandra} Deep Field North (CDF--N)}

A sample of 284 X--ray sources significantly detected in the hard band and inside 
an off-axis angle of 10 arcmin were selected from the 2 Ms CDF--N
source catalogue by Alexander et al. (\cite{Alexander2003}). The solid angle
area curve given by these authors was truncated at 10 arcmin off-axis angle as well.
For uniformity with the other samples used in this study,
the CDF--N fluxes were recomputed from the count rates 
given in Alexander et al., assuming a photon spectral index of 1.4 for all sources. 
To convert the count rates, which are extracted 
in the 2--8 keV band, into 2--10 keV fluxes, a counts to flux conversion factor of 
$2.88\cdot10^{-11}$ erg cm$^{-2}$ has been used.

Optical spectroscopic identifications in the CDF--N have been compiled 
from various catalogues in the literature. The largest number of spectroscopic
redshifts (114) were taken from Barger et al. (\cite{Barger2001,Barger2003}), and 
33 objects from Cohen et al., (\cite{Cohen2000}). Another 12 objects were taken
from various publications summarized in the NASA Extragalactic Database
(Bauer et al. \cite{Bauer2002}, Chapman et al. \cite{Chapman2003}, 
Cowie et al. \cite{Cowie2004}, Cristiani et al. \cite{Cristiani2004}, 
Smail et al. \cite{Smail2004}, Swinbank et al. \cite{Swinbank2004}, 
Wirth et al. \cite{Wirth2004}) and 6 objects were provided by P. Capak (2007,
priv. comm.).
The redshift of the object VLAJ123642+621331, originally placed at z=4.424 
by interpreting the single emission line in its spectrum as Ly$\alpha$
(Waddington et al. \cite{Waddington1999}, has been corrected to z=1.770, 
because the photometric redshift of this source (Capak 2007, priv. comm.) 
fits much better, if the line is interpreted as [OII]3727. 
With the set of 165 spectroscopic redshifts available, this CDF--N
subsample has a spectroscopic completeness of 58\%.
Photometric redshifts for the CDF--N have been kindly provided for this study
by P. Capak (2007, priv. comm.). They have been described to some degree in
Barger et al. (\cite{Barger2003}) and allow this field to be exploited to 
its full depth. Here 111 objects with photometric redshift errors $\Delta$z$<$1.5
have been used, leaving only 8 objects unidentified. The CDS--N sample utilized
here therefore has a redshift completeness of 97\%, very similar to the 
CDF--S discussed above. 

\begin{figure}[htp]
\begin{center}
\includegraphics[width=9truecm]{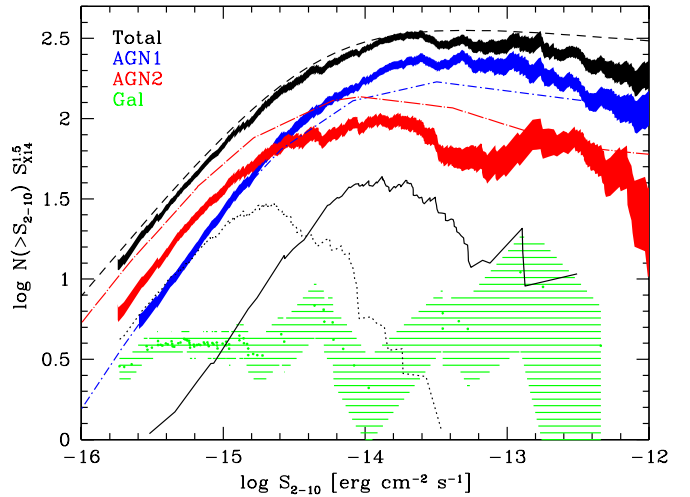}
\caption{Cumulative number counts for different subsamples of the data used here. 
For clarity the data have been normalized to a Euclidean slope 
(N($>$ S) $\cdot$ S$_{X14}^{1.5}$). The upper black band shows the data points 
and $1\sigma$ errors for the total sample in comparison to the analytical 
description (black dashed line) from Moretti et al. (\cite{Moretti2003}). The 
blue band shows the AGN type-1 number counts
in comparison to the model prediction (blue dot-short-dashed line) by GCH07. 
The red band is for type-2 AGN, again in comparison to 
the Gilli et al. prediction (red dot-long-dashed line). The green band and 
corresponding data points shows the number counts of galaxies (see text), which
are consistent with a 
Euclidean behaviour down to the faintest fluxes. The thin solid line peaking in 
the middle of the flux range shows the number counts for the $\sim 8\%$ 
unidentified sources in the sample and the thin dotted line the counts of those
sources with only photometric redshifts.}\label{fig:RNS}
\end{center}
\end{figure}
%

\begin{table*}[ht]
\begin{center}
\caption[]{Differential number counts best--fit parameters}
\begin{tabular}{@{}lccccc@{}}
\hline
Sample & N                & $\gamma_1$      & S$_b$/10$^{-14}$         & $\gamma_2$ & $\chi^2$ \\
       & [deg$^{-2}$]     &                 & [erg cm$^{-2}$ s$^{-1}$] &            &           \\
\hline
Total  & 167.9 $\pm$ 15.6 & 2.64 $\pm$ 0.03 & 0.99 $\pm$ 0.15   & 1.24 $\pm$ 0.06 & 1.35 \\
AGN1   & 122.2 $\pm$ 15.0 & 2.67 $\pm$ 0.03 & 1.39 $\pm$ 0.22   & 0.97 $\pm$ 0.08 & 1.36 \\
AGN2   &  35.6 $\pm$ 3.5  & 2.62 $\pm$ 0.04 & 0.33 $\pm$ 0.06   & 0.82 $\pm$ 0.15 & 1.36 \\
\hline
\end{tabular}\label{tab:nsfit}
\end{center}
\end{table*}

\section{AGN classification criteria}
\label{sec:class}

A crucial prerequisite for the analysis presented here is the discrimination
between absorbed and 
some controversy in the literature. The classical distinction between optically 
unobscured (type-1) and obscured (type-2) AGN is done using a discrimination 
through optical spectroscopy. Optical type-1 AGN have broad permitted emission 
lines ($>$2000 km/s), while optical type-2 AGN do not show broad permitted lines, but still have 
high-excitation narrow emission lines. Therefore many works in the field are mainly 
using the presence of broad lines as discriminator and classify objects as type-1 
AGN only if they have broad lines (see e.g. Barger et al. \cite{Barger2005}; 
Treister \& Urry \cite{Treister2006}). However, this AGN classification scheme breaks 
down, when the optical spectrum is insufficient to accurately determine the emission
line widths. At high redshifts and low luminosities there are several effects that 
compromise the optical classification. First, in general high redshift objects are 
faint, so that they require very long observing times on large telescopes 
to obtain spectra of sufficient 
quality allowing to unambiguously discern broad emission line components. Secondly, 
there are redshift ranges, where the strong classical broad lines shift out of the 
observed spectroscopic optical bands. Thirdly, and probably most importantly, at high 
redshifts the spectroscopic slit includes the light of the whole host galaxy, which
dilutes the spectrum of the AGN nucleus. 
Depending on the ratio between nuclear and host luminosity, the 
host galaxy can easily outshine the AGN nucleus, rendering the AGN invisible in the 
optical light. This effect has been demonstrated for local bona fide bright Seyfert
galaxies, where the emission lines are diluted in the host galaxy light when the 
integral light of the galaxy is sent through the spectrograph slit (see e.g. Moran 
et al. \cite{Moran2002}) and is the main reason, why X-ray samples are much more 
efficient in picking up low-luminosity AGN at high redshift. 

The optical spectroscopic AGN classification is 
mainly based on broad permitted lines in the restframe blue and ultraviolet part 
of the spectrum. The amount of obscuration in the optical spectrum can be estimated
using e.g. the Balmer decrement. At $N_{\rm H}>3 \cdot 10^{21}$ cm$^2$ the 
broad UV lines are usually suppressed and only narrow lines remain.    

In principle, a classification purely based on X-ray properties would be possible. 
Large hydrogen column densities block soft X--rays. Therefore, a suitable AGN 
classification scheme in X--rays could involve the column density $N_{\rm H}$ that 
can be determined from the X--ray spectra. Usually, faint AGN X--ray spectra are fit by 
simple power law models. The effect of increasing column density is that the soft 
part of the spectrum is more and more suppressed, i.e. the spectrum becomes harder 
at high $N_{\rm H}$. Many works in the field use an $N_{\rm H}$ value of $10^{22}$
to discriminate between X--ray absorbed and unabsorbed objects. 
However, in practice it is not possible to determine $N_{\rm H}$ values for samples with 
typically very faint X-ray sources, due to the small number of observed photons. 
Therefore, one has to resort to hardness ratios measured from coarse X-ray bands. 
At high redshifts in general it gets more difficult to determine absorption values
from X--ray spectral information, because the absorption cutoff is gradually shifting
out of the observed band to lower energies. In particular, there are several claims in the 
literature that at high redshifts BLAGN tend to show significant absorption (see e.g. Brusa 
et al., \cite{Brusa2003}; Wang et al., \cite{Wang2007}), but substantial systematic effects 
have to be taken into account here. First, N$_{\rm H}$ values can only be positive, so that 
the scatter in the data is producing spurious absorption values even for perfectly unobscured 
objects (see e.g. the discussion in the appendix of Tozzi et al., \cite{Tozzi2006}). Those 
spurious absorptions get larger with increasing redshift. Also, warm, ionized absorbers are 
present in many AGN, producing a deep trough in the X-ray spectrum around the oxygen edge 
($\sim 0.5 keV$), with a continuum recovering at softer energies. If the redshift of the 
object is such that the rest frame soft X-ray emission is shifted out of the observed band, 
these ionized absorbers can easily be mistaken as intrinsic cold gas absorption. This is in 
particular true for high-redshift Broad Absorption Line (BAL) quasars (see e.g. Hasinger et 
al., 2004). 

\begin{figure*}[htp]
\begin{center}
\includegraphics[width=14truecm]{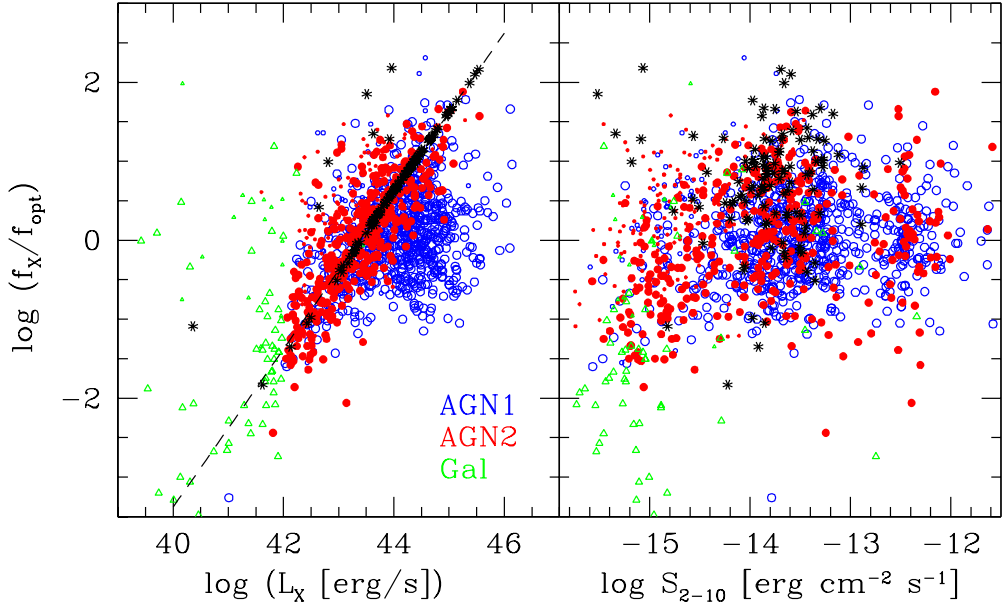}
\caption{The optical to X--ray flux ratios of the sample objects as a function 
of observed X--ray luminosity (left) and X--ray flux (right). Objects classified as type-1 AGN
are shown by blue open circles, type-2 AGN as filled red circles and galaxies
as green triangles. Objects with spectroscopic redshift information are plotted
with larger symbols and those, for which only photometric redshifts exist, with 
small symbols. The unidentified objects, for which crude redshifts 
have been assigned, are shown as black asterisks.
}\label{fig:fxo}
\end{center}
\end{figure*}

To overcome the difficulties of classification in either the optical or the X-ray bands, 
a combined optical/X--ray classification scheme has been introduced (Schmidt et al., 
\cite{Schmidt1998}, Szokoly et al. \cite{Szokoly2004}, Zheng et al. \cite{Zheng2004}). 
It first involves a threshold X--ray luminosity, which is set at 
$log(L_{\rm X})>42$ erg s$^{-1}$ to discriminate against the most luminous star forming 
galaxies. All point--like X--ray sources with
higher luminosities are considered as AGN. The second parameter is the X-ray hardness 
ratio HR, defined as 
\begin{equation}
{\rm HR} = ({\rm H}-{\rm S}) / ({\rm H}+{\rm S}),
\end{equation}
with S and H being the count rates in the soft and hard band, respectively. This is 
illustrated in Figure~\ref{fig:hrz}, which shows the observed hardness ratios
as a function of redshift for objects selected from XMM--Newton and Chandra surveys
with high--quality optical spectroscopy and well--determined hardness ratios.
The differences in the response curves of the different XMM--Newton and Chandra 
surveys are small enough
($\Delta HR \sim 0.1$) that they do not significantly distort the observed hardness ratio 
distributions.  The 
left part of the figure shows objects spectroscopically classified as broad--line AGN 
(BLAGN), while the right part shows AGN without broad emission lines.  
The BLAGN show a very narrow range of hardness ratios, independent of redshift.
The reason for this is that BLAGN are strongly dominated
by unabsorbed, powerlaw--type X--ray spectra, which do not change shape when they 
are redshifted. Only 11 out of 320 objects in this sample (3.4\%) have  
HR values larger than -0.2, which can therefore conveniently be used as a threshold 
discriminating between 
X--ray absorbed and  unabsorbed objects. BAL QSO are treated as BLAGN in this 
context. They are among the most absorbed BLAGN. On the contrary, the non-BLAGN show hardness 
ratios spread out over a much larger range. In particular there is a kind of bimodal 
distribution,  about half of the objects have HR values smaller than -0.2, consistent 
with unabsorbed spectra, and the rest have larger hardness ratios. There is also a
clear trend of decreasing hardness ratios with redshift, which is due to the fact
that at higher redshifts the exponential low--energy cutoff in the X--ray spectra
of absorbed sources is shifting towards lower energies. The latter effect leads
to a degeneracy between absorbed and unabsorbed X--ray sources at high redshifts, 
which can no longer be discriminated using their hardness ratios.

Given this information, both the optical and X--ray selection effects and systematic 
errors in the distinction between absorbed and unabsorbed sources can be largely 
overcome by combining optical and X--ray spectral diagnostics. For those sources, 
which are unambiguously classified either as BLAGN or as NLAGN, using their 
optical/NIR spectra, the optical type-1/type-2 classification is maintained. 
For simplicity, Seyfert sub-classes up to Sy1.5 (i.e. 1.2, 1.4 and 1.5) 
are sorted into the type-1 class, subclasses above the Sy1.5 (i.e 1.8, 1.9)
are sorted into the type-2 class. Sources, for which optical spectroscopy
does not allow an unambiguous discrimination between type-1 and type-2 are classified using 
their hardness ratios.  An AGN  with a hardness ratio ${\rm HR}<-0.2$ is called 
type-1, and with ${\rm HR}>-0.2$ type-2. This classification has also the advantage 
that it can be applied for sources with no spectroscopic information,
but which have reliable photometric redshifts. The other advantage is that it is based on a 
simple measurable quantity, the hardness ratio which can in principle be included 
in models of the X--ray background. In addition, there are also obvious systematic effects and 
difficulties. A small fraction of AGN are known not to follow the simple 
optical/X-ray obscuration/absorption correlation, either broad line AGN have absorbed 
X--ray spectra (e.g. for BAL QSOs, see e.g. Hasinger et al., \cite{Hasinger2002}; 
Chartas et al., \cite{Chartas2002}), or narrow-line AGN have apparently unabsorbed X-ray spectra 
(see e.g. Panessa \& Bassani, \cite{Panessa2002}). 
Also, as shown in Figure~\ref{fig:hrz}, the hardness ratio threshold corresponds to
different NH-values as a function of redshift. Finally, the dust obscuration  
affecting the optical properties of AGN versus gas absorption affecting the X--rays
may be different, depending on environment and redshift. Nevertheless,  
these systematic effects are arguably far smaller than those of any of the other 
classifications individually. It is, however, important to reiterate that the 
threshold hardness ratio of -0.2 used here is calibrated against the optical 
discrimination between BLAGN and non-BLAGN and therefore corresponds to a 
somewhat
lower N${\rm H}$ value ($\sim3\cdot10^{21}~cm^{-2}$) than that typically used in the literature. 

Figure~\ref{fig:hsam} shows the AGN sample utilized in this 
study in the optical magnitude--X--ray flux and the redshift--luminosity plane, 
respectively. Type--1 and type-2 objects, classified with the above procedure 
are indicated by filled and open circles. In principle it would be possible to
calculate emitted rest-frame luminosities from the known redshift and an absorption 
column density estimeted from the hardness ratios for each object separately. 
However, the small number of source photons detected for most objects in this 
sample introduces significant statistical errors in the results. Due to the positive 
bound on the N$_H$ values, systematic biases are introduced into the analysis.
Throughout this paper, therefore, only observed luminosities in the observed 2-10 keV band
are used. The effects of intrinsic rest-frame luminosities can always be estimated 
using a forward modelling technique like e.g. in the GCH07 population synthesis
model.

%
%

\section{Number Counts of different populations}
\label{sec:lns}

The combination of a large number of surveys with a wide range of sensitivity
limits and solid angle coverage presents a unique resource. On one hand, the 
surveys presented here resolve a large fraction of the 2--10 keV X--ray 
background. On the other hand, we have an almost complete optical identification 
and redshift determination for all components. We are thus in the position to 
study the contribution of different object classes to the X--ray background.
Using the solid angle versus flux limit curve given in Figure~\ref{fig:Effarea} 
the number counts for different classes of sources were constructed. 
Figure~\ref{fig:RNS} gives the cumulative source counts normalized to a 
Euclidean behaviour: N($>$S$_{2-10}$) S$_{\rm X14}^{1.5}$, where  S$_{\rm X14}$ is the 
2--10 keV flux in units of 10$^{-14}$ erg cm$^{-2}$ s$^{-1}$. 
A Euclidean relation would be a horizontal line in this graph. Separate 
curves are shown for the total sample, the type-1 and type-2 AGN, as well as 
for galaxies. The total and the AGN source counts display the well-known 
deviation from a Euclidean behaviour at low fluxes. For each of these populations
a smoothly broken power was fit to the differential source counts:
\begin{equation}
{\rm N(S) = (~(N S_{\rm X14}^{\gamma1})^{-1} + ( (N/{S_b^{\gamma2-\gamma1}} S_{\rm X14}^{\gamma2})^{-1}~)^{-1}} 
\end{equation}
The best--fit parameters for those are given in Table \ref{tab:nsfit}. 
As predicted in the GCH07 population synthesis 
model, type-1 AGN dominate at high fluxes, while type-2 AGN start to 
dominate at fluxes below $10^{-14}$ erg cm$^{-2}$ s$^{-1}$. There is, however,  
a significant difference between the observed number counts 
and the model prediction, which will be addressed later.

An earlier attempt to disentangle the contribution of the different source
classes to the number counts has been done by Bauer et al., \cite{Bauer2004} 
(see also Brandt \& Hasinger \cite{Brandt2005}).
However, there are systematic differences in this analysis: the 
redshift incompleteness and the different classification schemes 
for absorbed and unabsorbed AGN make a direct comparison difficult.
If we compare the surface densities for different classes of sources 
above a 2-10 keV flux of $2.5 \times 10^{-16}~erg~cm^{-2}~s^{-1}$, 
close to the sensitivity limit of the deep Chandra surveys, a very good match 
is found for the total X-ray source population ($\sim4400~deg^{-2}$ for both 
analyses). The type-1 AGN surface density derived in this analysis ($\sim 
1330~deg^{-2}$) is somewhat higher than the surface density of X-ray unabsorbed
objects in Bauer et al. ($\sim780~deg^{-2}$). On the other hand,
the type-2 AGN surface density here is correspondingly lower ($2250~deg^{-2}$ 
versus $\sim2900~deg^{-2}$), respectively.  
Figure~\ref{fig:RNS} also shows that galaxies, defined in this 
analysis as X--ray sources with luminosities smaller than 
10$^{42}$ erg cm$^{-2}$ s$^{-1}$, appear at the lowest fluxes and are 
consistent with a Euclidean behaviour. Their space density at a flux of 
 $2.5 \times 10^{-16}~erg~cm^{-2}~s^{-1}$ is $\sim 840~deg^{-2}$, compared
to the range of $370-640~deg^{-2}$ observed by Bauer et al. The number counts 
of the galaxies defined in this way are about a factor of 3 higher than those 
predicted by Ranalli et al. (\cite{Ranalli2003}) for purely star forming 
galaxies. This excess could indicate that the objects in this sample still have
a substantial AGN contribution. Some of these X-ray sources could also be 
associated to diffuse emission in groups of galaxies. 
The figure also shows that the unidentified sources 
are conveniently situated at intermediate fluxes, so that they are unlikely
to introduce a bias in the number counts. At fainter fluxes a larger proportion
of objects are optically too faint for reliable spectroscopy, so that
photometric redshifts have to be employed.

\begin{figure}[htp]
\begin{center}
\includegraphics[width=9truecm]{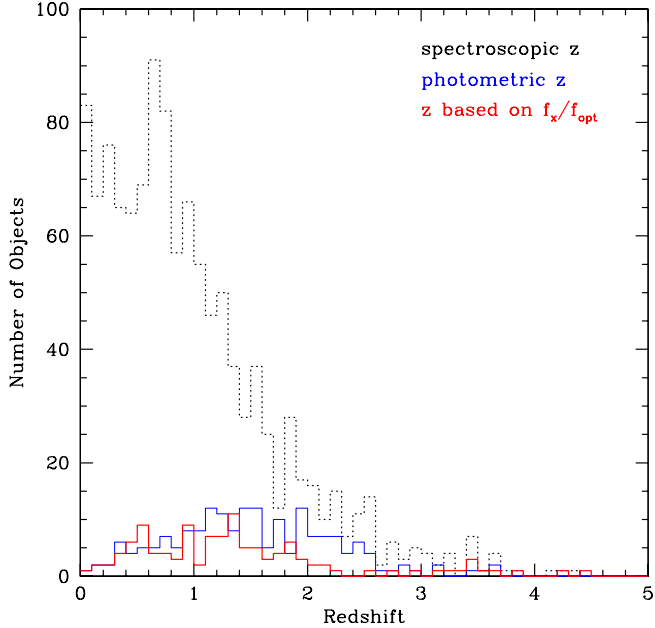}
\caption{Redshift distribution of objects with high-quality spectroscopic redshifts 
(thick black histogram), with photometric redshifts (thin blue histogram) and with crude redshifts
estimated using the X--ray to optical flux ratios (dashed red histogram).}\label{fig:redsh}
\end{center}
\end{figure}

\section{Correcting redshift incompleteness}
\label{sec:crude}

As noted in Table \ref{tab:samp}, about 8\% of the sample remains 
without redshift identification, compared to about 13\% with only photometric redshifts. 
While this is a small fraction, the incompleteness most likely depends on redshift,
so that systematic redshift and therefore spurious evolution effects may be introduced 
in the analysis. On the other hand, as Figure~\ref{fig:RNS} shows, the unidentified 
sources are mainly at intermediate fluxes. The reason is that they are typically from 
serendipitous {\em Chandra} and {\em XMM--Newton} surveys, where the quality of the 
optical photometry available does not allow a photometric redshift determination as
accurately as in the deepest fields, which have excellent multi--band photometry. 
 
It has been shown by Fiore et al. \cite{Fiore2003} that for objects, where the optical
light is dominated by their host galaxy (i.e. non Broad--line AGN), there exists
a correlation between X--ray luminosity and the optical to X--ray flux ratio 
f$_{\rm x}$/f$_{\rm opt}$. This is probably mainly due to the fact that AGN host galaxies 
have a rather limited range in absolute luminosity, while AGN luminosities can occupy a 
much larger range. In order to test this conjecture for the AGN sample utilized here, the
f$_{\rm x}$/f$_{\rm opt}$ ratios are plotted against luminosity and X--ray flux in 
Figure~\ref{fig:fxo}. As is well known, the X--ray sources show a substantial 
scatter in their X--yay to optical properties. However, unlike type-1 AGN, which 
at high luminosity are dominated by their non--thermal energy output both in the 
optical and X--ray regime, type-2 AGN do indeed show a significant trend of higher 
X--ray/optical flux ratios with increasing redshift. This correlation could be fit 
by a simple proportionality between X--ray luminosity and X--ray to optical flux ratio:
\begin{equation}
log(L_{\rm X}) = log ({\rm f}_{\rm X}/{\rm f}_{\rm opt}) + 44.38 
\end{equation}
The slope of this relation is the same as in the work of Fiore et al.
\cite{Fiore2003}. For all objects without redshift information a crude redshift 
was estimated using this relation, also for the objects in the HELLAS2XMM survey, where 
crude redshifts have already been estimated with a similar method by Fiore et al. 
(\cite{Fiore2003}) and Mignoli et al. (\cite{Mignoli2004}).
Bauer et al. (\cite{Bauer2004}) and Barger et al. (\cite{Barger2005}) have
shown that the Fiore et al. relation breaks down for very faint optical 
counterparts (R$>$25). Indeed, a few objects indicated with asterisks in Figure~\ref{fig:fxo},
originating in the CDF-S and CDF-N surveys 
fall off from the main correlation line because their optical counterparts are too faint.
For these objects, statistically rather uncertain photometric redhifts have been utilized
instead of the crude redshifts derived from the Fiore et al. relation.

Figure~\ref{fig:redsh} shows the redshift distribution, separately for  
spectroscopic, photometric and crude redshifts. It is reassuring that the 
distribution of the crude redshifts is very similar to that of the photometric 
redshifts. Given the fact that the flux distribution of the unidentified sources 
peaks in the middle of the sample range, the crude redshifts will likely not bias 
the results. Although individually they have a very low reliability, in a statistical
treatment they should give a fair estimate of the systematic effects involved in 
the redshift incompleteness. 

Since the AGN classification scheme discussed above, in the absence of optical
spectroscopy can still discriminate between type-1 and type-2 AGN, it is also
applied to the unidentified sources with crude redshifts in order to estimate
luminosities and to study the effect of the incompleteness on the analysis
discussed below.

\section{The fraction of absorbed AGN}
\subsection{Luminosity--dependence of type-2 fraction}

A comparison of the open and closed circles (AGN type-2 and type-1, respective
ly)
in the right panel of figure \ref{fig:hsam} shows that there is a much larger fr
action
of type-2 AGN at lower luminosities. Indeed, this confirms what had been
found previously by different teams. Ueda et al. (\cite{Ueda2003}) studied the h
ard X--ray luminosity
functions of $\sim$230 AGN, based largely on {\em ASCA} and deep {\em Chandra} s
urveys, with about
 95\% completeness. They found that the fraction of type-2 AGN decreases with X
--ray luminosity,
$L_{\rm 2-10\,keV}$. Similar results have been obtained independently from diffe
rent samples by
Steffen et al. (\cite{Steffen2003}) and Hasinger (\cite{Hasinger2004}) and later
by La Franca et al. (\cite{LaFranca2005}).

To demonstrate the trend observed in the sample studied here, the type-2 fraction (i.e. the number 
of type-2 AGN divided by the total number of AGN in any particular subsample) was first calculated
as a function of X--ray luminosity. In order to minimize possible redshift effects, the 
analysis was carried out in the redshift range $0.2\le z\le 3.2$. This 
redshift range was chosen for simplicity to be compatible with the redshift 
shells used in the later analysis and to minimize any systematic effects that
could happen at the lowest and highest redshifts. The results are shown
in Table \ref{tab:t2fl} and Figure~\ref{fig:t2f} (left). In order to keep track of systematic
selection effects introduced by the unidentified sources, the calculation was 
performed separately for the sample ignoring the unidentified sources (thin blue error 
bars in the figure and quantities with primes in the table) and for that with 
crude redshifts included (thick black symbols in the figure and quantities without primes in 
the table). Because the individual subsamples of type-1 and type-2 AGN sometimes 
contain small numbers, for which Gaussian statistics do not apply, the Gaussian equivalent error bars 
of the type-2 fraction were computed following Gehrels ({\cite{Gehrels1986}): $\Delta N=\sqrt{N+0.75}$.

\begin{table}[ht]
\begin{center}
\caption[]{Type--2 fraction as a function of luminosity, measured in the redshift interval z=0.2--3.2}
\begin{tabular}{@{}cccccccc@{}}

\hline
Lmin  &  Lmax  &  t1'  &  t2'  &   t2'/(t1'+t2') &  t1  &  t2  &   t2/(t1+t2) \\
\hline
42.0  &  42.2  &   4  &  18  &  0.82$\pm$0.08  &    5  &  18  &  0.78$\pm$0.08\\
42.2  &  42.5  &   8  &  40  &  0.83$\pm$0.05  &    9  &  40  &  0.82$\pm$0.06\\
42.5  &  42.7  &  24  &  38  &  0.61$\pm$0.06  &   24  &  39  &  0.63$\pm$0.06\\
42.7  &  43.0  &  25  &  52  &  0.68$\pm$0.05  &   26  &  53  &  0.67$\pm$0.05\\
43.0  &  43.2  &  38  &  40  &  0.51$\pm$0.06  &   41  &  42  &  0.51$\pm$0.06\\
43.2  &  43.5  &  48  &  52  &  0.52$\pm$0.05  &   54  &  53  &  0.50$\pm$0.05\\
43.5  &  43.7  &  58  &  65  &  0.53$\pm$0.05  &   65  &  72  &  0.53$\pm$0.04\\
43.7  &  44.0  &  71  &  54  &  0.43$\pm$0.04  &   71  &  64  &  0.47$\pm$0.04\\
44.0  &  44.2  &  93  &  31  &  0.25$\pm$0.04  &  103  &  41  &  0.28$\pm$0.04\\
44.2  &  44.5  & 121  &  20  &  0.14$\pm$0.03  &  128  &  34  &  0.21$\pm$0.03\\
44.5  &  44.7  &  88  &  14  &  0.14$\pm$0.03  &   92  &  21  &  0.19$\pm$0.04\\
44.7  &  45.0  &  58  &  12  &  0.17$\pm$0.05  &   58  &  17  &  0.23$\pm$0.05\\
45.0  &  45.2  &  25  &   3  &  0.11$\pm$0.06  &   26  &   6  &  0.19$\pm$0.07\\
45.2  &  45.5  &  16  &   1  &  0.06$\pm$0.06  &   16  &   1  &  0.06$\pm$0.06\\
45.5  &  46.2  &  13  &   1  &  0.07$\pm$0.07  &   13  &   1  &  0.07$\pm$0.07\\
\hline
\end{tabular}\label{tab:t2fl}
\end{center}
{\small Quantities t1' and t2' with primes refer to the sample ignoring the unidentified sources, while quantities t1, t2 without primes refer to the sample including crude redshifts for the unidentified sources.}
\end{table}

\begin{figure*}[htp]
\begin{center}
\includegraphics[width=9truecm,clip=]{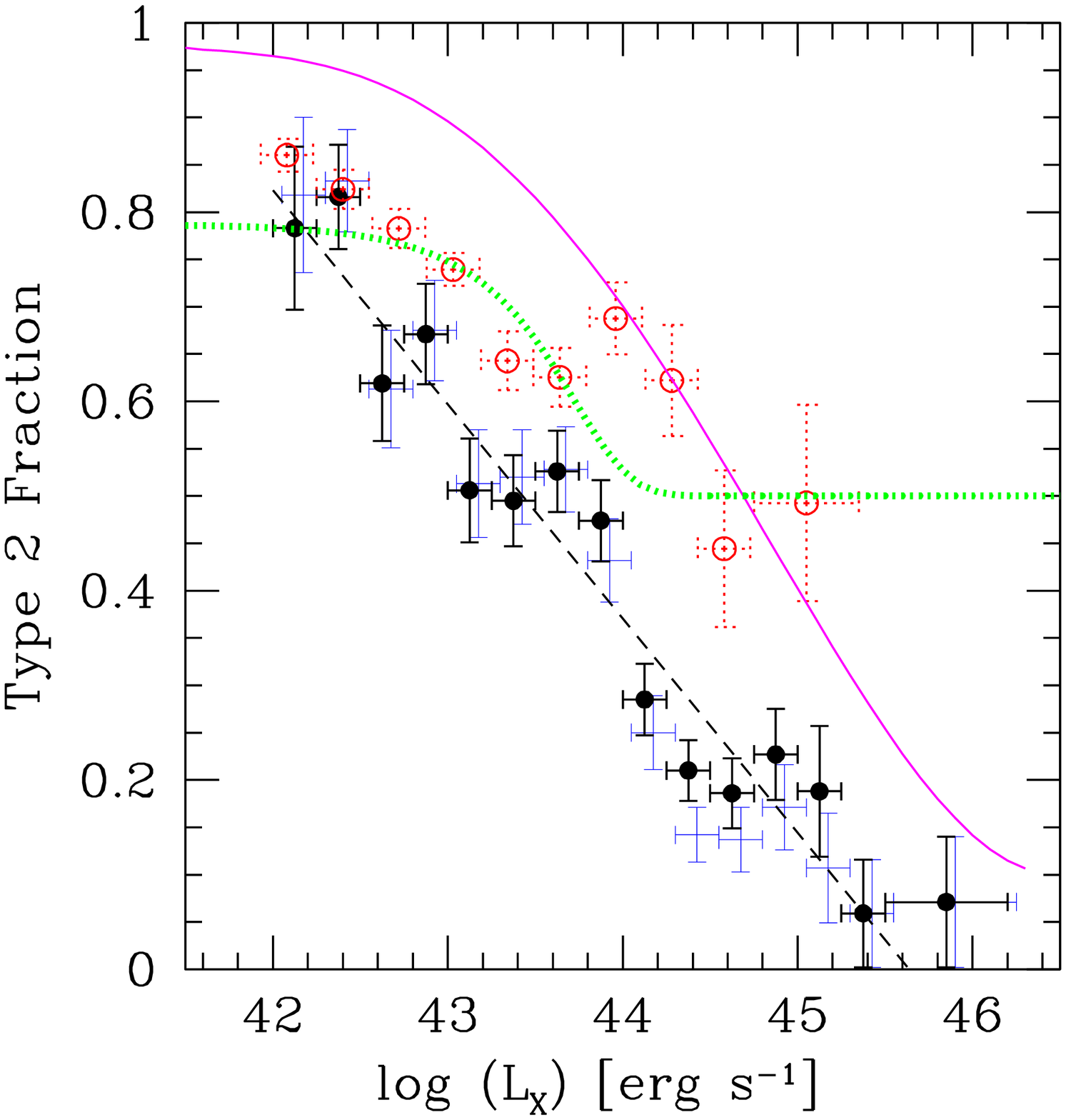}
\includegraphics[width=9truecm]{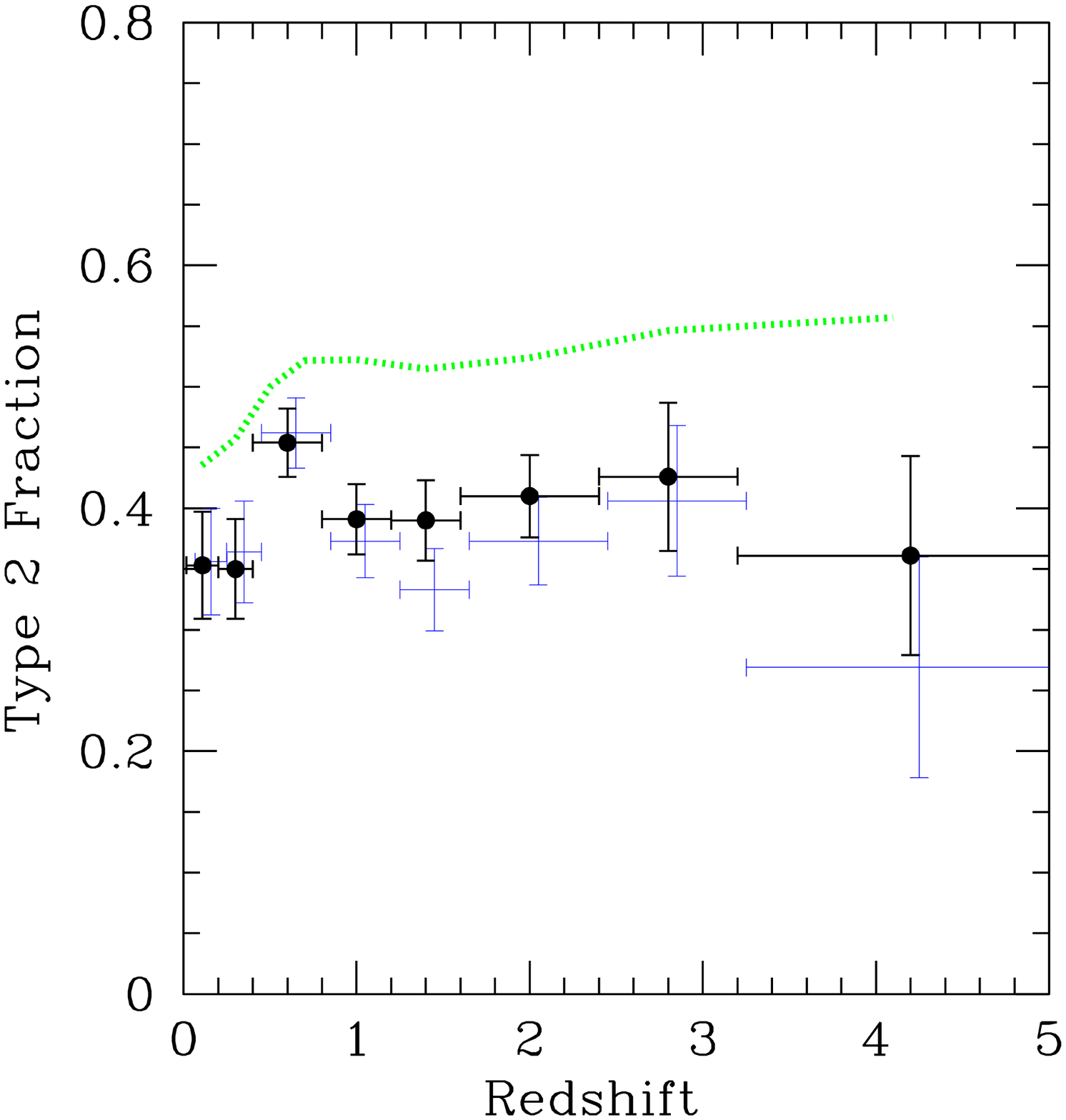}
\caption{Left: Type--2 fraction as a function of luminosity, based on the data in the redshift 
range z=0.2--3.2. The thin blue data points show the sample ignoring the 
unidentified sources, while the thick black data points with filled circles 
include the crude redshifts derived in Section \ref{sec:crude}. 
For clarity the thin blue points are shifted slightly to higher luminosity.
A simple linear
function fit to the black data points is shown as a dashed black line. 
The data points with red circles are derived using optical line widths 
of a sample of [OIII] selected AGN from the SDSS (Simpson \cite{Simpson2005}). 
The solid magenta line is from silicate dust studies of a sample of AGN observed with Spitzer 
in the MIR (Maiolino et al. \cite{Maiolino2007}). The dotted green line shows the ratio of 
Compton-thin absorbed AGN to all Compton-thin AGN assumed in 
the GCH07 population synthesis model.
Right: The observed fraction of type-2 AGN as a function of redshift. 
The colour coding of the thin blue and thick black data points is the same 
as in the left figure. The green solid line shows the prediction 
of the GCH07 model, assuming no redshift evolution in the absorbed fraction
and folding over the solid angle sensitivity curve from Figure \ref{fig:Effarea}. 
The error bars along the Y-axis in this and later figures all give 1-$\sigma$
uncertainties.
}\label{fig:t2f}
\end{center}
\end{figure*}

Figure~\ref{fig:t2f} (left) shows the trend of a decreasing absorption fraction with 
increasing X--ray luminosity with unprecedented accuracy. 
This figure also demonstrates that the redshift incompleteness effects are 
moderate and mainly affect objects at higher luminosity. A simple linear function 
has been fit to the completeness--corrected data points. It yields a slope of $-0.226\pm0.014$
and a marginally acceptable reduced $\chi^2$ of 1.96. A flattening 
of the decline is anyhow expected at the highest luminosities, because the simple linear
fit does not preclude negative values. A comparison of this new data with 
other results in the literature will be made in Section \ref{sec:disc}. 

\subsection{Redshift--dependence of AGN type-2 fraction}

In order to address a possible evolution of the obscuration fraction with 
redshift, one can first simply determine the observed ratio of type-2 versus 
total AGN, integrated over all luminosities, in shells of increasing redshift. 
This is very similar to what was done by Treister \& Urry (\cite{Treister2006}). Figure 
\ref{fig:t2f} (right) shows the dependence of the observed fraction of absorbed
sources on redshift. To first order this data indicates a flat behaviour, apart 
from a small increase at redshifts below z=0.8. The diagram also shows the 
type-2 fraction as a function of redshift predicted for the current sample 
from the most recent background synthesis model (GCH07), which assumes no 
evolution in the absorbed fraction. This curve shows that the small rise of 
the observed type-2 fraction at the lowest redshifts can be understood as an 
effect of the different flux limits for type-1 and type-2 sources in each 
survey (the latter ones are harder to find because of absorption). The 
comparison with this model and with the Treister \& Urry data points will be 
discussed in Section \ref{sec:disc}.

\begin{figure*}[htp]
\begin{center}
\includegraphics[width=12truecm,clip=]{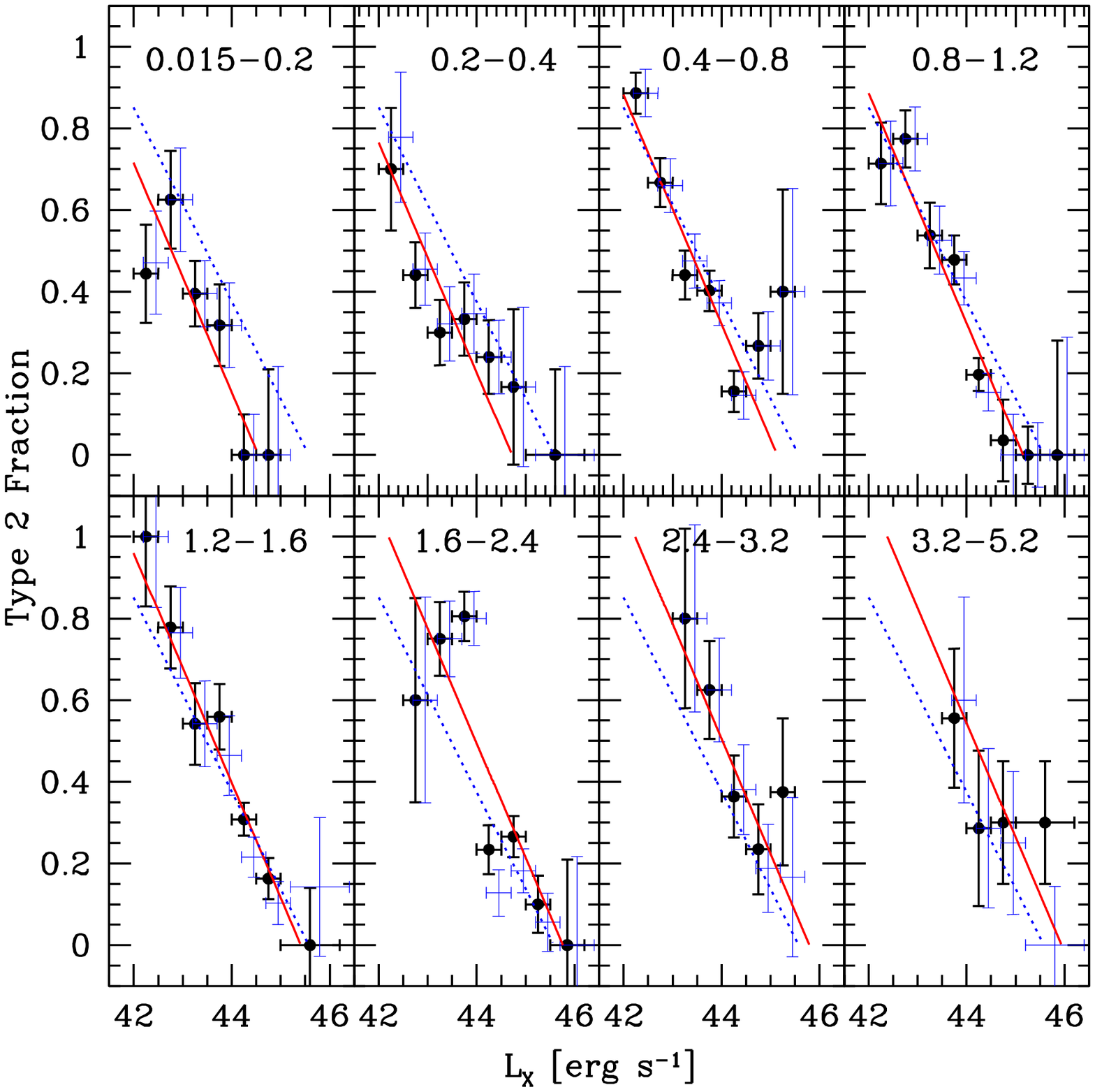}
\caption{Variation of the type 2 fraction with luminosity for different redshift
shells. The thin blue data points show the sample ignoring the
unidentified sources, while the thick black data points with filled circles
include the crude redshifts derived in Section \ref{sec:crude}. 
The dotted blue line gives the fit to the average relation 
in the redshift range 0.2--3.2. 
}\label{fig:t2flz}
\end{center}
\end{figure*}

\begin{table}[ht]
\begin{center}
\caption[]{Type--2 fraction norm in redshift intervals}
\begin{tabular}{@{}lcccc@{}}
\hline
\hline
z Range    &   Norm               &  corr     & corr. Norm       & Slope \\
\hline
0.015-0.2  &   0.224 $\pm$   0.045  &   0.83    &0.269  $\pm$  0.054 & -0.25 $\pm$ 0.06\\
0.2-0.4    &   0.273 $\pm$   0.041  &   0.87    &0.313  $\pm$  0.047 & -0.16 $\pm$ 0.05\\
0.4-0.8    &   0.390 $\pm$   0.025  &   0.95    &0.409  $\pm$  0.026 & -0.28 $\pm$ 0.03\\
0.8-1.2    &   0.394 $\pm$   0.027  &   1.00    &0.396  $\pm$  0.027 & -0.29 $\pm$ 0.03\\
1.2-1.6    &   0.468 $\pm$   0.030  &   0.98    &0.476  $\pm$  0.031 & -0.30 $\pm$ 0.04\\
1.6-2.4    &   0.564 $\pm$   0.030  &   1.00    &0.565  $\pm$  0.030 & -0.36 $\pm$ 0.05\\
2.4-3.2    &   0.575 $\pm$   0.060  &   1.04    &0.551  $\pm$  0.058 & -0.27 $\pm$ 0.11\\
3.2-5.2    &   0.615 $\pm$   0.086  &   1.06    &0.578  $\pm$  0.081 & -0.11 $\pm$ 0.12\\
\hline
\end{tabular}\label{tab:fitzcorr}
\end{center}
\end{table}

Figure~\ref{fig:t2f} (right) is integrated over all luminosities in each 
redshift 
shell and is therefore possibly hiding trends, which might be observable in a 
better resolved parameter space. For a better diagnostic of the redshift 
evolution of the type-2 fraction, the dependence of the type-2 fraction on 
luminosity was analysed separately in different redshift shells. In 
Figure~\ref{fig:t2flz} the same trend of a decreasing absorption fraction
as a function of X--ray luminosity, as observed in the total sample (see 
Figure~\ref{fig:t2f} left) is confirmed in each individual redshift shell. The
linear fit to the total sample in the redshift range 0.2--3.2 from Figure 
\ref{fig:t2f} left) is shown
as a dotted blue line in each panel. The same linear trend has been fit to the 
data points in each individual panel. First both the normalization of the 
curve at $log(L_X)=43.75$ as well as the slope of the relation have been left 
as free parameters in the individual fits. The last column of Table \ref{tab:fitzcorr}
shows the slopes derived for the individual fits. Since they are all consistent 
with each other, the average slope over all panels has been determined to 
$-0.281\pm0.016$. s slope is somewhat steeper than that derived for the total
sample in the redshift range 0.2-3.2 (see above), which is a consequence of the 
increasing absorbed fraction with redshift.

The data in the individual panels have then been again fit with a linear 
relation, but this time keeping the slope fixed to the average value of -0.281.
The results of these fits are shown as the solid red lines in 
Figure~\ref{fig:t2flz} and also in Table \ref{tab:fitzcorr}. For every 
redshift shell this table shows in column (2) the norm of the linear fit to the 
luminosity-dependence, taken at a value of $log(L_X)=43.75$ in the middle of 
the range.  There is a clear trend of an increasing normalization of the 
absorbed fraction with redshift which can be seen in the comparison of the blue
dotted and red solid lines in Figure~\ref{fig:t2flz} and in the corresponding 
data in the table. However, in order to quantitatively estimate the evolution 
of the absorbed fraction with redshift, the systematic selection effects have 
to be corrected for. They are due to the fact that in every flux--limited sample
it is harder to detect absorbed objects compared to unabsorbed ones of the same
intrinsic flux. Folding this selection effect over the solid angle as a function
of flux limit of the current meta-sample, one can in principle obtain a 
correction curve as a function of redshift. For simplicity, here the green 
curve in Figure~\ref{fig:t2f} (right) determined for the GCH07 model  
folded with the effective solid angle sensitivity curve in Figure
\ref{fig:Effarea} and normalized to a redshift of 2 has been used. The (small)
correction values are given in column (3) of Table \ref{tab:fitzcorr}. With this 
correction applied, finally the corrected normalization can be calculated for 
every redshift shell, wich is given as column (4) of Table \ref{tab:fitzcorr}.

This corrected normalization has been plotted as a function of redshift in 
Figure \ref{fig:norm}. A significant increase of the absorbed fraction norm 
with redshift is seen from this figure. This increase, however, seems to 
saturate above a redshift of z$\sim$2. Therefore, a broken power law has been 
fit to the data: $t_2(z)=t_2(0)\cdot (1+z)^{\alpha_1}$ for $z \le z_b$ , 
saturating at $t_2(z) = t_2(z_b)$ for $z>z_b$.  
The best-fit parameters are determined to $t_2(0)=0.279 \pm 0.024$,
$\alpha_1=0.62\pm0.11$, $z_b=2.06\pm0.47$ and the reduced $\chi^2$ of this fit 
is 0.82. The evolution in the absorbed fraction in the redshift range up to 
z$\sim$2 is therefore confirmed with a statistical significance of $5.7\sigma$. 
In order to assess the significance of the apparent evolution saturation at high
redshifts and to compare with previous works, also a simple power law
$ t_2(z) = t_2(0)\cdot (1+z)^{\alpha_2}$ has been fit to the whole redshift 
range. The best-fit parameters for this representation are 
$t_2(0)=0.308 \pm 0.022$ and $\alpha_2=0.48\pm0.08$ (i.e. a significance of 
$6.1\sigma$). With a reduced $\chi^2$ of 1.30 this fit is worse than for the 
broken power law model but still marginally acceptable.  Finally, a constant 
value can be ruled out with very high confidence (reduced $\chi^2=6.37$).

A simple power law $ t_2(z) = t_2(0)\cdot (1+z)^{\alpha_3}$ has also been 
fit to the data without including the crude redshifts (thin blue data points in
all diagrams). The best-fit parameters for this data set are 
$t_2(0)=0.309 \pm 0.024$ and $\alpha_3=0.38\pm0.09$ (a significance of 
$4.4\sigma$) with a reduced $\chi^2$ of 1.08. Including the crude redshift thus
pronounces the evolution in the redshift range up to z=2 and provides evidence 
for a saturation of this evolution. In order to assess the significance of this
saturation it is interesting to note the actual numbers of objects in the last 
two redshift bins, both including and excluding the crude redshifts. The 
redshift bin $2.4<z\leq3.2$ contains 39 (38) type-1 and 29 (26) type-2 AGN 
(numbers in brackets refer to the sample without crude redshifts). The redshift
bin $3.2<z\leq5.2$ contains 23 (19) type-1 AGN and 13 (7) type-2 AGN. The 
inclusion of crude redshifts is thus affecting mostly the highest redshif
bin.

\begin{figure}[htp]
\begin{center}
\includegraphics[width=9truecm,clip=]{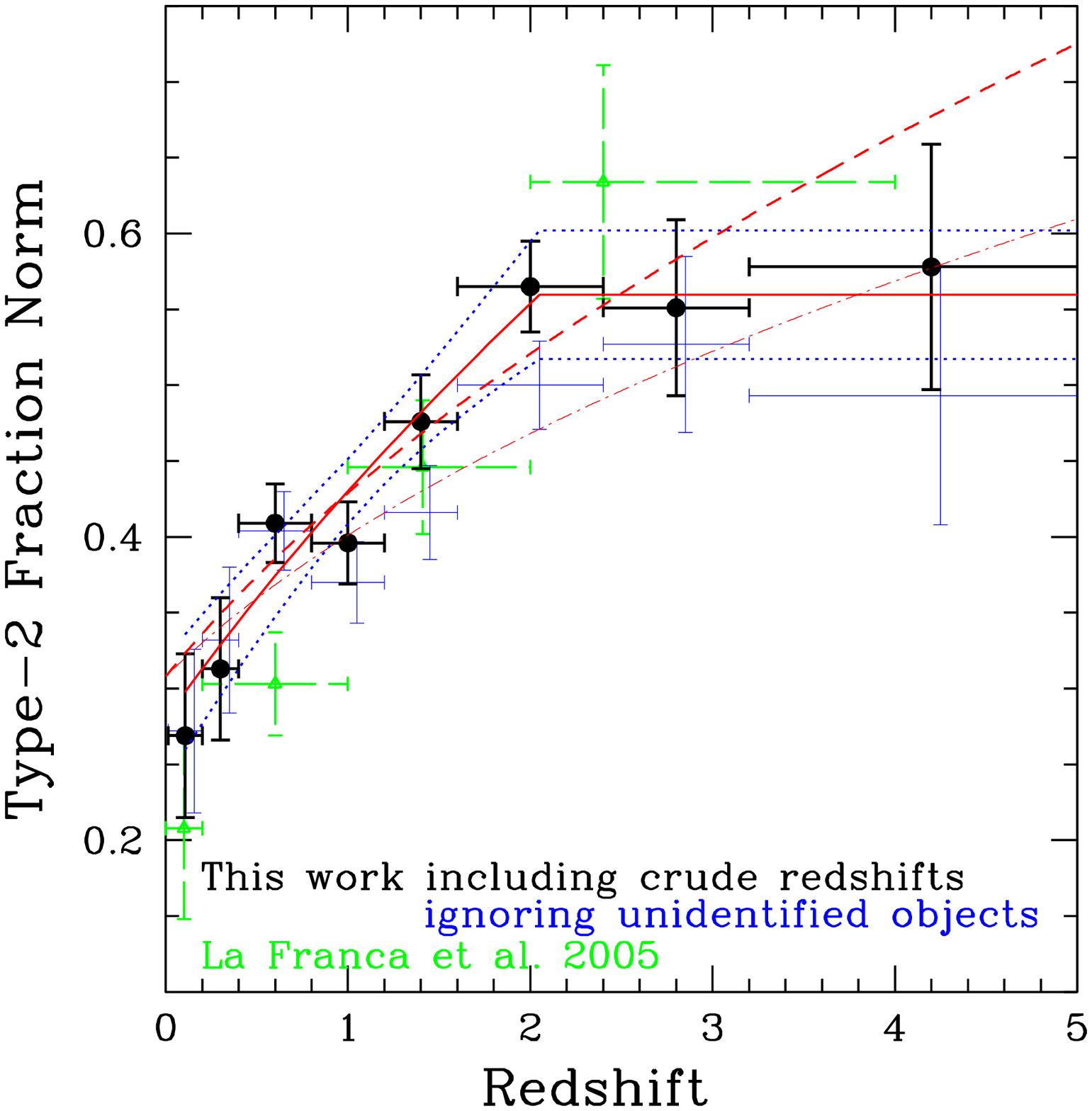}
\caption{
Dependence of the normalization of the type-2 fraction at log(L$_{\rm X}$)=43.75
with redshift (from the fits shown in Figure~\ref{fig:t2flz}). 
The red solid curve shows a saturated power law fit to the solid black data (see text).
The blue dotted lines give the 90\% confidence intervals for this fit.
The dashed red line shows a single poer law fit to the black solid data 
over the full redshift range, while the dot-dashed line shows the same single power 
law fir to the thin blue data points without including the crude redshifts. 
The dashed green data points are from La Franca et al. (\cite{LaFranca2005}).
}\label{fig:norm}
\end{center}
\end{figure}

\section{Discussion}
\label{sec:disc}

\subsection{Systematic effects of different AGN classification schemes}
\label{sec:sys}

In section \ref{sec:class} different classification techniques for X--ray selected AGN have been discussed. In most classical treatments, type-1 AGN are just defined by the presence of broad permitted lines observed in their optical spectra (see e.g. Steffen et al., \cite{Steffen2003}; Barger et al., \cite{Barger2005}; Treister \& Urry \cite{Treister2006}). In order to estimate the systematic differences with the combined optical/X--ray classification employed in this paper, the luminosity- and redshift dependence of the type-2 fraction has been recomputed for the sample using the same BLAGN classification
scheme. In this case type-1 AGN are spectroscopically identified 
objects with broad emission lines, while type-2 AGN are all other
spectroscopically identified objects. For consistency with other works, 
objects with only photometric or crude redshifts are ignored here.   
Figure~\ref{fig:blagn} shows the results of this analysis in comparison to the  combined X-ray/optical classification and to the work from Treister \& Urry (\cite{Treister2006}) and Treister et al. (\cite{Treister2008}), respectively.
The left panel shows the observed dependence of the type-2 fraction on X-ray luminosity. The data points derived from the BLAGN only classification (green) are fully consistent with the data from Treister et al. (\cite{Treister2008}), but at lower X--ray luminosities both are significantly higher than those from the combined X-ray/optical classification used here. This is mainly due to the large number of unabsorbed AGN at low to intermediate luminosities, missing in the BLAGN sample presumably due to dilution from the host galaxy light (see Figure~\ref{fig:hrz}). The same trend is also seen in the right panel of Figure
\ref{fig:blagn}, which shows a significant difference in the redshift 
dependence of the the absorbed fractions determined by the two different 
methods.  

\begin{figure*}[htp]
\begin{center}
\includegraphics[width=9truecm]{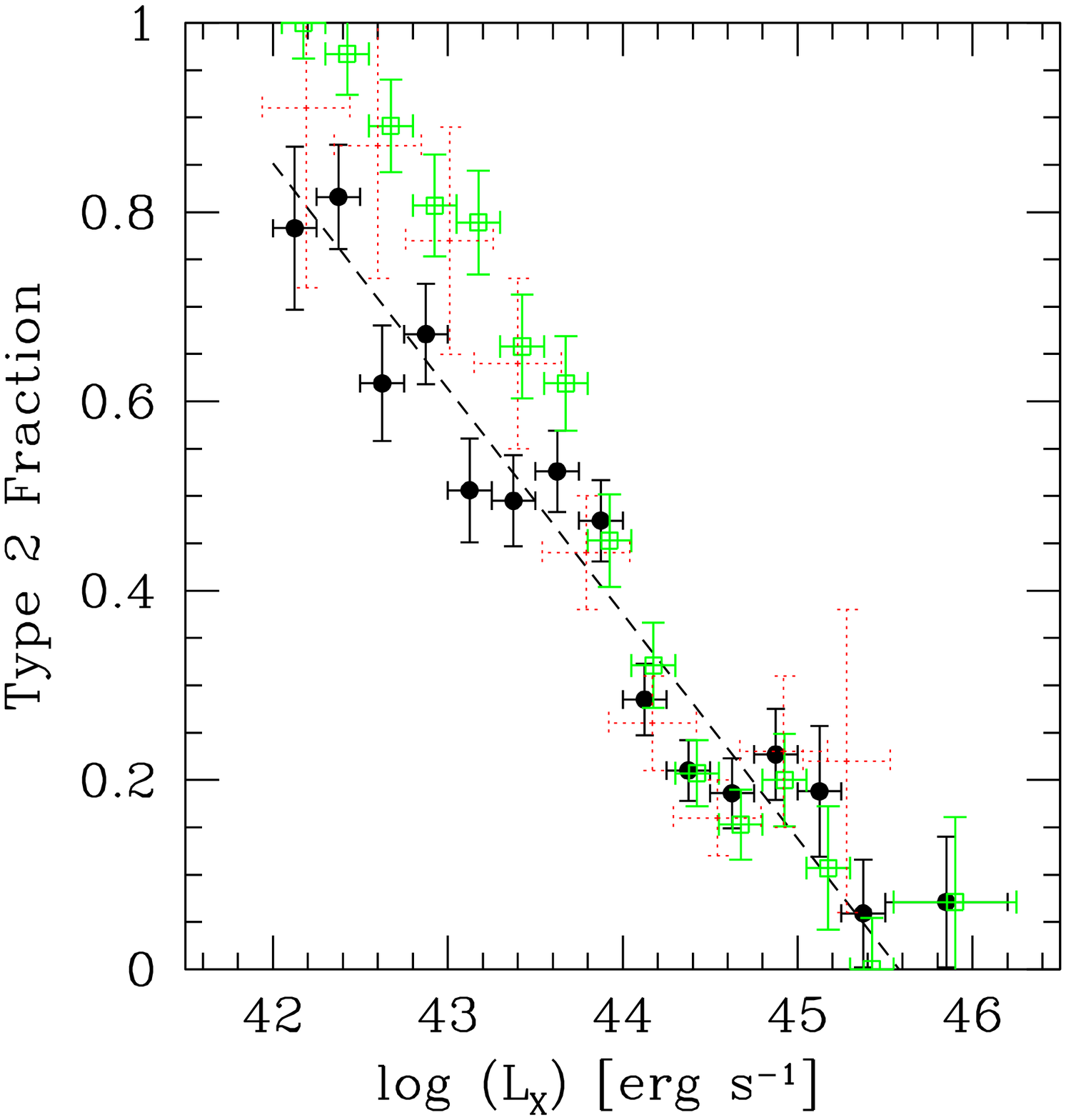}
\includegraphics[width=9truecm]{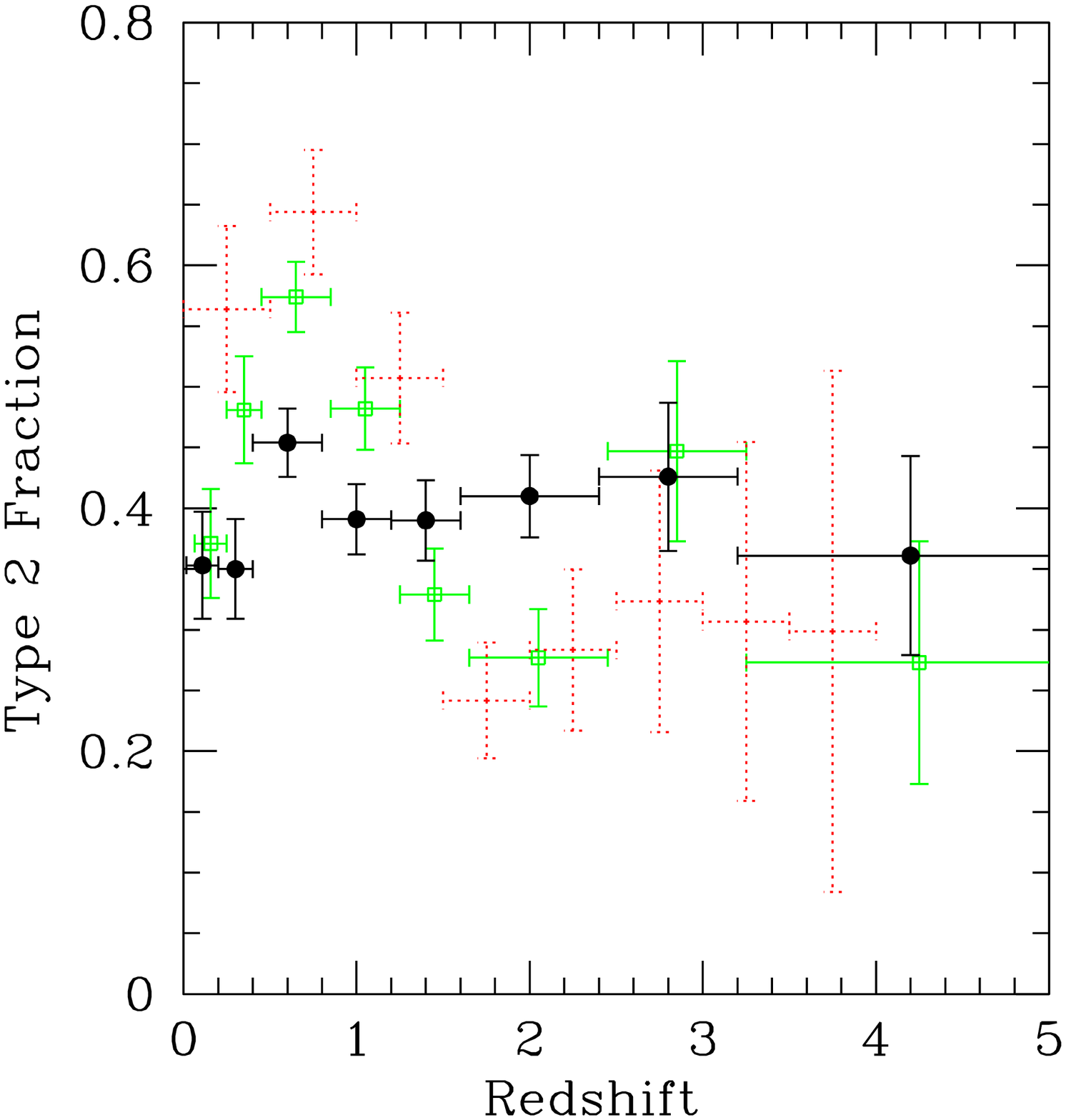}
\caption{Comparison of type-2 fractions as a function of luminosity (left) and a function of redshift 
(right) for different AGN classification schemes. The filled black data points with thick error bars 
correspond to the combined optical/X--ray classification introduced in section \ref{sec:class} and 
are identical to those in Figure~\ref{fig:t2f}. The green data points with open 
squares and thin error bars have been derived using a purely optical spectroscopic classification, 
where only BLAGN are classified as type-1 objects (see Figure~\ref{fig:hrz}). Red data points with dotted
error bars are from Treister et al. 
(\cite{Treister2008}) (left) and from Treister \& Urry \cite{Treister2006} (right), respectively.}\label{fig:blagn}
\end{center}
\end{figure*}

\subsection{Luminosity dependence of absorbed fraction}

All analyses represented in Figure~\ref{fig:t2f} (left), based on X-ray and optical samples, 
agree that there is a strong decline of the absorbed fraction with AGN luminosity. The X--ray 
analysis presented here confirms this trend in much more detail and with better statistical 
and systematic errors than previous X--ray analyses. Similar trends are found in other 
wavebands. Local Seyfert galaxies selected from the SDSS spectroscopy sample using the [OIII] 
lines show a fraction of BLAGN increasing with the luminosity of the isotropically emitted 
[O III] narrow emission line (Simpson \cite{Simpson2005}). 
In order to convert the [O III] line luminosity into a 2-10 keV luminosity 
for the display in Figure~\ref{fig:t2f}, the 
empirical luminosity-dependent correlations found by Netzer et al. (\cite{Netzer2006}) for 
type-1 and type-2 Seyfert galaxies were averaged: 

\begin{equation}
\log{  \frac{ {\rm L}_{[OIII]}}{ {\rm L}_{2-10} } } = 13.7 -0.358 \log{ {\rm L}_{2-10} }.
\end{equation}

Mid--infrared (MIR) Spitzer low resolution spectra of 25 luminous QSOS at redshifts between 
2 and 3.5, combined with low--luminosity type-1 AGN confirm this picture (Maiolino et al. 
\cite{Maiolino2007}). These authors find a strong, non--linear decrease (a factor of ten) of 
the thermal continuum dust emission at 6.7$\mu$m with increasing 5100 \AA~optical continuum 
emission and [OIII]5007 line luminosity. They interprete this as a decreasing covering factor 
of circumnuclear dust as a function of luminosity. They clearly detect silicate dust emission 
features (see also Sturm et al., \cite{Sturm2006}), whose strength correlates with luminosity, 
accretion rate $L/L_{\rm Edd}$ and black hole mass $M_{\rm BH}$. 

Independently, Treister et al. (\cite{Treister2008}) recently attempted to measure this covering fraction by studying the ratio of the MIR to bolometric luminosity in a sample of z$\sim$1 BLAGN selected optically in the Sloan Digital Sky Survey (SDSS), the Great Observatories Origins Deep Survey (GOODS) and the Cosmic Evolution Survey (COSMOS). They use the Spitzer 24$\mu$ flux as a proxy of the MIR dust-reprocessed radiation.  For their redshift range this corresponds to $\sim 12\mu$, which according to Maiolino et al. (\cite{Maiolino2007}) and Sturm et al. (\cite{Sturm2005}) is mostly dominated by the silicate emission and less due to the reprocessed thermal dust emission. The silicate emission is thought to come mostly from the narrow-line region (NLR) and thus does not directly probe the covering factor of the molecular dust torus. To estimate the bolometric luminosity of their AGN (primarily the blue bump) they use the Galex NUV data, sampling the rest frame waveband 875-1400\AA, which is possibley affected by absorption from the Ly-forest and even small amounts of dust extinction. Given these systematic differences, it is surprising that their results almost perfectly match the Maiolino et al. curve in Figure~\ref{fig:t2f} (left).
 
Of course, all the sample selections represented in Figure~\ref{fig:t2f} (left)
are subject to systematic errors and selection biases. X--ray surveys in the
2--10 keV band almost completely miss the population of Compton--thick Seyfert
galaxies, which are barely detectable only above 10 keV (see e.g. GCH07). 
However, first significant minority samples of Compton-thick candidate objects 
have been detected in deep-wide X-ray surveys (Tozzi et al., \cite{Tozzi2006}; 
Hasinger et al., \cite{Hasinger2007}; Brunner et al.,  \cite{Brunner2008}), 
consistent with the GCH07 model predictions. Including information from deep 
MIR surveys, there is now convincing evidence that the population of 
Compton--thick intermediate--luminosity AGN even at higher redshifts is of the 
same order as that of Compton-thin AGN (see e.g. the recent results in the CDFS
by Daddi et al. \cite{Daddi2007} and Fiore et al. \cite{Fiore2008a} as well as 
in the COSMOS field by Fiore et al. \cite{Fiore2008b}), again fully consistent 
with the predictions of the GCH07 model. Optical surveys, on the other hand, 
while including many of the Compton-thick AGN missing in X--ray samples, 
fail on
the AGN populations diluted by their host galaxy (see Figure 
\ref{fig:hrz}), and also those so heavily obscured that very little scattering
or ionizing radiation escapes (see e.g. Ueda et al., \cite{Ueda2007}). 
Nevertheless, the agreement of better than 20\% between the optical sample 
selected from the SDSS using the [OIII] lines (Simpson \cite{Simpson2005}) and 
the hard X--ray selection from this paper is reassuring in this respect. 

In Figure~\ref{fig:t2f} also the relation between absorbed
Compton-thin AGN to all Compton-thin AGN assumed in the GCH07 X--ray background
population synthesis model is shown. Because
Compton-thick objects have been explicitely excluded in this curve, it 
can be
directly compared to the observed 2-10 keV X-ray sample results. Please note,
however that this is the assumed intrinsic ratio, the real observations are 
still affected by additional selection effects. Nevertheless, it is apparent 
that this model overpredicts the fraction of obscured sources, in particular at 
high luminosities. This is also likely one of the reasons for the systematic 
difference between the predicted and observed number counts for type-1 and 
type-2 AGN in Figure~\ref{fig:RNS}. Probably the best approximation of the 
"true" fraction of obscured objects is the Maiolino et al. relation. Hopefully,
the GCH07 model can be updated accordingly in the future.

Physically, this luminosity--dependent fraction might be interpreted as a 
'cleanout effect' because more luminous AGN can dissociate, ionize and finally
blow away the dust in their environment. At face value this trend signals a 
break--down of the simple unified AGN model, where the difference between 
type-1 and type-2 AGN is solely in the geometry of the observer's line of sight
with respect to the absorber (Antonucci \cite{Antonucci1993}). The luminosity 
dependence of the absorbed fraction
indicates an intrinsic physical difference in the average absorber properties 
as a function of intrinsic AGN luminosity. The simplest interpretation of these
findings is that low-luminosity AGN are surrounded by an absorbing/obscuring 
medium which covers a large solid angle ($\sim 80\%$ of the sky as seen from 
the black hole), however with an average column density much lower 
($N_H\sim10^{22}~cm^{-2}$) than the putative Compton-thick dusty torus of the 
unified model ($N_H\sim10^{24-25}~cm^{-2}$). High-luminosity AGN are then able 
to clean out their environment, either by ionizing the surrounding medium, or 
by blowing it away through an outflowing wind (see also M\"uller \& Hasinger 
\cite{Mueller2007}). This behaviour is also a key ingredient in the 
"blast wave AGN feedback" model presented very recently by Menci et al. 
(\cite{Menci2008}), where AGN accretion luminosity, feedback and absorption 
are intimately connected to each other.  

\subsection{Redshift dependence of absorbed fraction}

The evolution of the luminosity--dependent obscuration fraction with redshift 
is still a matter of debate. A possible redshift dependence of the obscured 
fraction was reported by La Franca et al. (\cite{LaFranca2005}), using the 
HELLAS2XMM sample combined with other published catalogs, based on a total of 
508 AGNs selected in the 2--10 keV band. On the other hand, other authors (e.g.
Ueda et al. \cite{Ueda2003}, GCH07) did not find a significant redshift 
dependence. Recently, Treister and Urry (\cite{Treister2006}) 
performed an analysis, using an AGN meta--sample of 2300 AGN with $\sim50\%$ 
redshift completeness and find a shallow increase of the type-2 AGN fraction 
proportional to $(1+z)^{0.4}$. The differences between the various results 
are very likely due to the different sample sizes and the way, in which the 
redshift incompleteness, the different flux limits and the AGN classification 
have been taken into account in the analysis. 

The analysis in this paper shows 
a significant trend of an increasing absorbed fraction with redshift. 
Interestingly, at first sight, the observed fraction of absorbed sources, 
after correction for the identification incompleteness has a 
redshift dependence very similar to the prediction of the 
no evolution GCH07 model 
(see Figure~\ref{fig:t2f}, right). This, however, turns 
out to be a "conspiracy" due to a combination of the shape of the luminosity-- 
and redshift--dependence of the absorbed fraction. As can be seen in 
Figure~\ref{fig:hsam}, at higher redshifts the median of the sample shifts 
to higher luminosities. In the case of the almost linear decrease of the 
absorbed fraction with luminosity observed in Figure~\ref{fig:t2f} (left) one 
would expect a significant decrease of the observed absorbed fraction if 
there was no redshift evolution (see also the model predictions in 
Figure 2 of the Treister \& Urry \cite{Treister2006} paper). This has 
to be compensated by an increase of the absorbed fraction with redshift
if an almost constant behaviour like the one in Figure~\ref{fig:t2f} (right) 
is observed.
Alternatively, if at high luminosities the absorbed fraction is constant, 
as assumed in the GCH07 model, an almost constant trend with redshift is
expected, assuming no evolution.

However, disentangling the data into separate redshift and luminosity
intervals resolves the degeneracy of the analyses based on one dimensional
analyses. 
Figure~\ref{fig:t2flz} confirms with high statistical accuracy that the 
decreasing trend of the absorbed fraction with luminosity is present in all 
redshift intervals. A similar result has already been shown in broader redshift
and luminosity bins by Barger et al. (\cite{Barger2005}), however, their results
are affected by the systematic differences discussed in subsection 
\ref{sec:sys}.  
However, the Figure~\ref{fig:t2flz} also shows an increase of the absorbed fraction
with redshift. If the data in each redshift slice are fit by a simple linear 
relation, the normalization of this relation, here assumed for a luminosity 
log(L$_{\rm{X}})$ = 43.75 in the middle of the observed range, clearly shows
an increase with redshift up to z$\sim$2, where this trend seems to saturate. Figure
\ref{fig:norm} shows that this trend is highly significant and can be fit by a
power law evolution proporional to (1+z)$^\alpha$, with $\alpha=0.62\pm0.11$. 
This evolution is thus substantially faster than the exponent $\alpha=0.4\pm0.1$
originally found by Treister \& Urry (\cite{Treister2006}). Several reasons can 
be envisaged for this difference:
First, these authors have fit the power law trend over the full 
redshift range, because the flattening of the curve was not apparent in 
their data. In addition they used the BLAGN classification only, which 
thends to over--estimate the absorbed fraction at low redshifts (see Figure 
\ref{fig:blagn}). Finally, the inclusion of crude redshifts in the analysis 
presented here tends to recover more absorbed objects at higher redshifts and higher
luminosities. 

Interestingly, the increase of the normalization of the absorbed fraction 
relation with redshift is very similar to the observed fractions given in 
figure 6b of the the La Franca et al. (\cite{LaFranca2005}) paper based 
mainly on the HELLAS2XMM sample (see also Figure~\ref{fig:norm}). This 
could, however, be somewhat of a chance coincidence, because their data 
shows the observed ratios integrated over a broad luminosity range
(more like Figure~\ref{fig:t2f} (right) in this paper), and has its own 
systematic
selection effects. Nevertheless, the trend originally claimed by 
La Franca et al. (\cite{LaFranca2005}) is fully confirmed here.  

It is also interesting to note that the saturated absorption evolution
found in this study shows a redshift dependence very similar to the 
space density evolution of the overall AGN population with an increase
from zero redshift to a peak redshift, where the evolution is saturating.
However, the absorption evolution is much shallower than the AGN density
evolution.

\subsection{Consequences for the broader context of AGN and galaxy co-evolution}

The cosmological evolution of AGN in the X--ray, optical and radio wavebends 
can be described by a luminosity-dependent density evolution model, in 
which the peak of the AGN space density shifts to lower luminosities 
towards lower redshifts (see above and Hopkins et al. \cite{Hopkins2007}). 
This evolutionary behaviour is very similar to the "cosmic downsizing" observed
in the bolometric luminosity of the normal galaxy population (Cowie et al.
\cite{Cowie1996}) and indicates that star formation and nuclear activity in 
galaxies go hand in hand. This kind of "anti-hierarchical" Black Hole growth 
scenario is not predicted by most semi-analytic AGN evolution models
based on galaxy merger scenarios (e.g. Kauffmann \& Haehnelt 
\cite{Kauffmann2000}; Wyithe \& Loeb \cite{Wyithe2003}; Croton et al. 
\cite{Croton2005}; Menci et al. \cite{Menci2008}; Rhook \& Haehnelt 
\cite{Rhook2008}). Typically these models are able to explain the space density
evolution of high--luminosity AGN, peaking at redshifts z$>$2, but have 
difficulties to match the behaviour at lower luminosities. Some authors (e.g. 
Treister et al. \cite{Treister2004}; Menci et al. \cite{Menci2008}) attribute
these differences to systematic selection biases against the identification of 
distant obscured, low-luminosity AGN, in particular if there is a strong 
evolution of absorption with redshift. The results of the analysis presented in 
this paper, although indeed confirming a significant evolution of obscuration,
argue against this conjecture, because the evolution observed across the 
whole redshift range is too shallow (at most a factor of 2) to explain
the dramatic luminosity-dependent density evolution effect. Therefore other
ingredients are required to explain the observed AGN downsizing. 

\begin{figure}[htp]
\begin{center}
\includegraphics[width=9truecm,clip=]{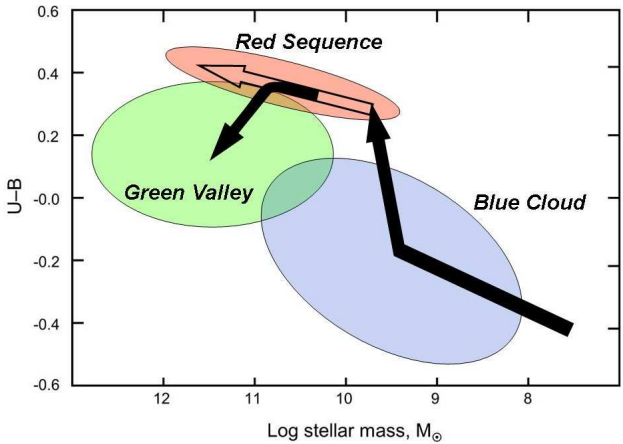}
\caption{
Schematic arrows showing galaxies moving in the colour-mass diagram under 
different evolutionary assumptions (following Faber et al.,
\cite{Faber2007}). It is conventionally assumed that galaxies 
on the "red sequence" are formed when two "blue cloud" galaxies merge 
with each other and the feedback from the growing central black hole
is quenching star formation, leaving a "dead" and red bulge-dominated
galaxy with a dormant supermassive black hole. Host galaxies of 
intermediate-luminosity active galactic nuclei are often found in the 
region between the blue cloud and red sequence, the so called "green valley".
The scenario described here tries to explain this by a "re--juvenation"
of bulge--dominated galaxies through the accretion of fresh gas from 
their environment. 
}\label{fig:GreenValley}
\end{center}
\end{figure}

Important inputs to the solution of this puzzle may come from the 
study of colours and morphology of the host galaxies of moderate-luminosity
AGN at low to intermediate redshifts. The SDSS in combination with data 
from the Galaxy Evolution Explorer (GALEX) satellite are of help here.
Kauffmann et al. (\cite{Kauffmann2007}) analyzed a volume-limited sample of 
massive bulge-dominated galaxies with data from both the SDSS and GALEX 
at redshifts 0.03$<$z$<$0.07. The GALEX NUV data are a very sensitive indicator
for low levels of star formation in these systems. These authors find that
practically all AGN in their sample have rather blue NUV-r colours and the UV 
excess 
light is almost always associated with an extended blue disk, while galaxies
with red outer regions almost never have a young bulge or a strong AGN. 
They suggest a scenario in which the gas of the outer disk is accreted
onto the bulge-dominated galaxy over cosmic time and provides the mass 
reservoir to trigger sporadic AGN accretion and star formation through 
gravitational instabilities (see also Cavaliere \& Vittorini \cite{Cavaliere2000}; Hammer et al. \cite{Hammer2005}). 
Interestingly, Schiminovich et al. (\cite{Schiminovich2007}) show that in the 
NUV-r versus mass diagram, the local SDSS AGN are preferentially at high masses,
but at colours in between the blue cloud and the red sequence, the so called 
"green valley" (see also Figure~\ref{fig:GreenValley}).

Similar results are found at higher redshifts from optical studies in deep 
and wide Chandra and XMM-Newton survey fields. Nandra et al. (\cite{Nandra2007})
discuss the colour-luminosity diagram for AGN
selected from the Chandra survey of the Extended Groth Strip in the redshift
range 0.6$<$z$<$1.4. They also find the AGN host galaxies in a distinct region 
of the colour-magnitude diagram, often in the "green valley". They interprete 
this in the context of star formation 
quenching in massive galaxies, resulting in a migration from the blue cloud to 
the red sequence. 
Silverman et al. (\cite{Silverman2008}) present an analysis of 109 
moderate-luminosity (log L$_{\rm X}<$42) AGN in the Extended Chandra Deep 
Field-South survey (Lehmer at al. \cite{Lehmer2005}) with redshifts 
z=0.4--1.1 and find a strong dependence of AGN host colours on the 
environment. Compared to all galaxies in the field they find that the 
fraction of galaxies hosting an AGN peaks in the "green valley", in particular
due to enhanced AGN activity in two narrow redshift spikes around z=0.63 
and z=0.73 (see Gilli et al. \cite{Gilli2005}). They also find that AGN 
host galaxies in this redshift range typically have hybrid morpholgies 
between pure bulges and disks. They try to explain their findings in an 
AGN merger scenario, but remark that the merger timescales are typically 
too short to explain the abundance of green valley galaxies. 

Putting these findings together, one arrives at a scenario possibly 
explaining the AGN cosmic downsizing as well as their colour and morphological 
evolution (see Figure~\ref{fig:GreenValley}). In order to produce an 
active galactic nucleus, we need two ingredients: a central supermassive 
black hole and efficient gas accretion.
The fuel can either be provided by a major merger driving gas into the 
center and simultaneously leading to an Eddington-limited growth of the 
black hole(s) (see e.g. Di Matteo et al. \cite{DiMatteo2005}; Li et al. 
\cite{Li2007}). In this case a bulge galaxy is formed and the feedback of the 
growing black hole can quench star formation. The result is a red 
bulge-dominated galaxy with a dormant supermassive black hole. The merger
time scale is, however, rather short ($\sim10^8$ years), so that only a
very small fraction of the galaxy population can be found in a transient
state between the blue cloud and the red sequence. It is likely that the 
space density evolution of the most luminous QSOs, peaking in the redshift 
range z=2-3, is a direct consequence of the merger history of massive halos.
The results above show that there is, however, also another source of 
gas supply, namely the accretion of fresh gas from the surrounding of a 
galaxy. In this way, a bulge-dominated galaxy with a central black hole 
which was formed during a previous merger event can be "re-juvenated", leading
to a disk surrounding the bulge. In the colour-mass diagram such a galaxy 
is therefore coming down from the red sequence into the green valley 
(see Figure~\ref{fig:GreenValley}) and
the galaxy assumes the hybrid bulge/disk morphology observed for the 
SDSS AGN and many of the X--ray selected intermediate-redshift AGN. The black
hole can be fed through episodic star formation and accretion events,
likely produced by gravitational instabilities in the outer disk. Since 
the black hole and the bulge have already been formed previously and the 
host galaxy is already relatively massive, the accretion typically happens 
at a rather low Eddington ratio, substantially smaller than in the 
merger case (see e.g. Marconi et al. \cite{Marconi2004}; Merloni \& Heinz
\cite{Merloni2008}). This mode of 
black hole growth is therefore slow enough that it can be observed in a
large fraction of all galaxies, providing a natural explanation for the 
late evolution in the ubiquitous lower-luminosity AGN. This scenario is
still rather descriptive and crude, but hopefully can be quantified by
including the bi-modal gas supply into semi-analytical evolution models 
in the future.

\section{Conclusions}
 
   \begin{enumerate}
      \item A meta--sample of hard X--ray selected AGN has been compiled from 
            the literature, providing an unprecedented combination of 
            statistical quality and spectroscopic as well as photometric 
            redshift completeness. Systematic differences in the 
            selection of this rather heterogeneous set of samples were
            taken care of as far as possible and should not significantly
            bias the results. 

      \item The small, but significant redshift incompleteness
            could be corrected for using the correlation between luminosity
            and X--ray to optical flux ratio observed in non-broad-line 
            AGN. A comparison between the analysis with and without the 
            inclusion of crude redshifts carried throughout the paper
            shows that the crude redshifts tend to enhance the high-redshift
            high-luminosity bins, but the main results do not depend 
            strongly on this choice. 

      \item The X--ray selected AGN could be classified using both 
            optical spectroscopy and X--ray hardness ratios. This 
            classification scheme appears more robust than either optical
            or X--ray classification alone. 

      \item A strong decrease of the fraction of absorbed AGN was found 
            with redshift. This trend confirms similar results obtained
            previously in the X-ray, but also optical and MIR bands. 
            As expected, systematic selection effects are present between
            different bands, but do not dominate the results. This 
            signals a break--down of the strong unified AGN model and
            indicates that high--luminosity AGN can clean out their 
            environment.

      \item The same decreasing trend with luminosity is found in all
            individual redshift shells up to z$\sim$3. However, the
            data indicate a significant evolution of the absorbed AGN
            fraction, which increases by about a factor of 2 up to 
            z$\sim$2 and remains approximately constant thereafter.

      \item This rather shallow evolution implies that the 
            luminosity--dependent AGN density evolution observed in 
            the X--ray, optical and radio bands is a real property of the 
            parent population and not caused by systematic selection 
            effects.

     \item  A new scenario is proposed, in which the cosmic down-sizing of 
            the AGN population is due to two different fuelling mechanisms,
            on one hand the efficient growth of luminous QSOs by galaxy 
            mergers early in the universe, on the other hand the re-juvenation
            of pre-formed bulges and black holes by slow gas accretion
            over cosmic time.
             
   \end{enumerate}

\begin{acknowledgements}
      I am indebted to Takamitsu Miyaji and Keisuke Shinozaki for providing
      the effective area curve, optical magnitudes and identification details
      of the HEAO--1/Grossan sample. I also am grateful to Yoshihiro Ueda for 
      providing me with the updated ASCA LSS and MSS hard X--ray samples and 
      effective area curves. Marcella
      Brusa and Fabrizio Fiore helped me to obtain some detailed unpublished
      information, e.g. X--ray hardness ratios and some new identifications
      for the HELLAS2XMM sample. I am particularly thankful to Peter Capak, who 
      provided me with the unpublished photometric redshifts from his PhD thesis
      for the CDF--N sources. 
      I thank John Silverman, Vincenzo Mainieri, Jaqueline Bergeron, Peter 
      Capak and Jeyhan Kartaltepe for the permission to use unpublished
      redshifts in the CDF--S. I am grateful to Roberto Della Ceca for providing
      me with unpublished details of the HBSS sample. I thank Roberto Gilli,
      Andrea Comastri, Marcella Brusa, Fabrizio Fiore, Nico Capelluti and 
      Roberto Maiolino for
      very helpful discussions and in particular Roberto Gilli for providing the
      model prediction in 
      Figure~\ref{fig:t2f} (right). Part of this work was supported by 
      the German \emph{Deut\-sche For\-schungs\-ge\-mein\-schaft, DFG\/} 
      Leibniz Prize (FKZ HA 1850/28--1). I thank the Institute for Astronomy,
      Manoa, Hawaii for the hospitality during my sabbatical visit in summer
      2007, when a significant part of this work was done.
      This research has made extensive use of the NASA/IPAC Extragalactic
      Database (NED) which is operated by the Jet Propulsion Laboratory,
      California Institute of Technology, under contract with the National
      Aeronautics and Space Administration, as well as the SIMBAD database,
      operated at CDS, Strasbourg, France. This work has made use of
      observations carried out using the Very Large Telescope at the ESO Paranal
      Observatory under Program ID(s): 170.A-0788, 074.A-0709, and 275.A-5060. 
      I thank a competent referee for very helpful and constructive 
      comments, which helped to improve the paper.

\end{acknowledgements}

{}

\end{document}